\documentclass[letter,preprint,aps,superscriptaddress,nofootinbib,tightenlines,preprintnumbers,cites]{revtex4}
\usepackage{graphicx}
\usepackage{enumerate}
\usepackage{bm}
\usepackage{slashed}
\usepackage{multirow}
\usepackage{verbatim}
\usepackage{amsmath,amssymb}
\usepackage{epsfig}
\usepackage{hhline}
\usepackage{placeins}

\newcommand\Eq[1]{Eq.~(\ref{eq:#1})}

\newcommand\Fig[1]{Fig.~\ref{fig:#1}}

\newcommand\Figs[2]{Figs.~\ref{fig:#1}-\ref{fig:#2}}
\newcommand\Sec[1]{Sec.~\ref{sec:#1}}

\newcommand\Tab[1]{Table~\ref{tab:#1}}
\newcommand{\be}{\begin{equation}}
\newcommand{\ee}{\end{equation}}
\newcommand\beq{\begin{eqnarray}}
\newcommand\eeq{\end{eqnarray}}

\newcommand{\mybar}[1]%
        {\kern 0.6pt\overline{\kern -0.6pt#1\kern -0.6pt}\kern 0.6pt}
\begin{document}

\preprint{MIT-CTP/4729}

\title{Low energy scattering phase shifts for meson-baryon systems}

\author{William Detmold}
\email{wdetmold@mit.edu}
\affiliation{Center for Theoretical Physics, Massachusetts Institute of Technology, Cambridge MA 02139, USA}

\author{Amy N. Nicholson}
 \email{anicholson@berkeley.edu} 
\affiliation{Department of Physics,
University of California, Berkeley, Berkeley CA 94720, USA}


\date{\today}
 
\begin{abstract}
In this work, we calculate meson-baryon scattering phase shifts in four channels using lattice QCD methods. From a set of calculations at four volumes, corresponding to spatial sizes of 2, 2.5, 3, and 4 fm, and a pion mass of $m_{\pi} \sim 390$ MeV, we determine the scattering lengths and effective ranges for these systems at the corresponding quark masses. We also perform the calculation at a lighter quark mass, $m_{\pi} \sim 230$ MeV, on the largest volume. Using these determinations, along with those in previous work, we perform a chiral extrapolation of the scattering lengths to the physical point after correcting for the effective range contributions using the multi-volume calculations performed at $m_{\pi} \sim 390$ MeV. 
\end{abstract}

\maketitle

\section{Introduction}

The study of hadronic interactions from first principles is one of the most exciting enterprises made possible by lattice QCD (LQCD) methods. In recent years, advances in algorithms and machines have allowed for high-precision calculation of meson-meson scattering phase shifts by several groups, including coupled channels and channels containing a resonance \cite{Wilson:2014cna,Wilson:2015dqa,Dudek:2012gj,Dudek:2012xn,Dudek:2014qha,Beane:2011sc,Aoki:2007rd,Aoki:2011yj,Pelissier:2012pi,Feng:2010es,Torres:2014vna,Bolton:2015psa,Briceno:2015dca,Lang:2012sv,Prelovsek:2013ela,Lang:2014yfa,Lang:2015hza,Lang:2011mn}. Progress for interactions between nucleons (\cite{Beane:2006mx,Beane:2011iw,Beane:2012vq,Detmold:2015daa,Beane:2015yha,Chang:2015qxa,Orginos:2015aya,Yamazaki:2012hi,Yamazaki:2015asa,Berkowitz:2015eaa,Murano:2013xxa}) has been much slower, due in part to the asymptotically exponential degradation of signal-to-noise for LQCD calculations involving nucleons \cite{Lepage:1989hd}. An important additional hindrance is the potentially poor convergence of Chiral Perturbation Theory ($\chi$PT) in the nucleonic sector, which is necessary for extrapolating current LQCD calculations, which are performed at unphysical quark masses, to the physical point. 

In this work, we study scattering of a single baryon and a meson. Using these systems, we can explore some questions of the convergence of $\chi$PT in a simpler setting without confronting excessive difficulties with statistical noise. While the scattering parameters for two baryon systems are constrained somewhat by chiral symmetry, having at least one meson in an initial or final state greatly simplifies the form of the expansion. Furthermore, the fine-tuning of interactions that exists in the nucleon-nucleon sector, leading to anomalously large scattering lengths and the need for non-perturbative treatment of the effective field theory \cite{Weinberg:1990rz,Weinberg:1991um,Kaplan:1996xu,Kaplan:1998tg}, is not expected to exist for the meson-baryon systems that we investigate.

Meson-baryon scattering is also of intrinsic interest for several reasons. Pion-nucleon and kaon-nucleon interactions are important in the determination of the equation of state of dense matter, particularly at densities relevant for neutron stars. Furthermore, meson-baryon interactions may be of interest for indirect reasons, such as understanding the final state interactions of various decays of interest for Standard Model phenomenology, disentangling single particle excited baryon states from meson-baryon states in LQCD calculations, or for understanding thermal contributions to nucleonic correlators in LQCD \cite{Tiburzi:2009zp,Tiburzi:2015tta,Bar:2015zwa}.

Experimental input exists for pion-nucleon scattering \cite{Schroder:1999uq,Schroder:2001rc}, as well as model-dependent extractions for kaon-nucleon scattering \cite{Martin:1980qe}. There is no direct experimental data for meson-baryon scattering processes involving hyperons. For an extensive discussion of experimental input to the relevant $\chi$PT analyses see \cite{Liu:2006xja,Liu:2007ct,Kaiser:2001hr,Mai:2009ce}. Additionally, there have been recent analyses of pion-nucleon scattering using Roy-Steiner equations \cite{Hoferichter:2015tha,Hoferichter:2015hva}. Two quenched LQCD calculations of meson-nucleon scattering processes have been performed \cite{Fukugita:1994ve,Meng:2003gm}, as well as dynamical two-flavor calculations of pion-nucleon scattering in the negative parity channel \cite{Verduci:2014csa,Lang:2012db}. Finally, a dynamical three-flavor, mixed-action result has been produced by the NPLQCD collaboration for the systems studied in this work \cite{Torok:2009dg}. These results were calculated at several values of the quark masses, and an extrapolation of the scattering lengths to the physical point was performed. However, these calculations made use of a single, relatively small volume, $(2.5 \, \mathrm{ fm})^3$, so that it is possible that they contain sizeable effective range corrections. Another LQCD calculation of the energies of many mesons and a single baryon using a larger volume, $(4 \, \mathrm{ fm})^3$, was performed in \cite{Detmold:2013gua}, and scattering lengths for the meson-baryon systems were also extracted from this data. These scattering lengths were found to differ significantly from the previous results on the smaller volume, suggesting that range corrections may indeed be large. 

In this work, we have calculated the low-energy scattering phase shifts of the four meson-baryon systems presented in \Tab{systems}. These systems are chosen to avoid the calculation of annihilation diagrams, which involve at least one $q\bar{q}$ same-flavor pair at source or sink, and which are computationally prohibitive at present. We perform calculations at a pion mass of $m_{\pi}\sim 400$ MeV at four volumes, $(2 \, \mathrm{fm})^3$, $(2.5 \, \mathrm{ fm})^3$, $(3 \, \mathrm{ fm})^3$, and $(4 \, \mathrm{ fm})^3$, and extract scattering lengths and effective ranges from this data. We find that the effective range contributions to the phase shifts on the smaller volumes are significantly larger than naively expected. We also perform these calculations at $m_{\pi}\sim 230$ MeV on the largest volume. We then use the data, combined with the data from Ref.~\cite{Torok:2009dg}, to perform a chiral extrapolation of the scattering lengths to the physical point, taking into account effective range contributions that are determined by the multi-volume study at $m_{\pi} \sim 390$ MeV.

This paper is organized as follows: in \Sec{lattice}, we discuss the details of our lattice calculation, and the methods we use to extract finite volume energy levels from correlation functions. Next, we discuss the calculation of scattering phase shifts from these energy levels in \Sec{phaseshifts}, and present our results for the phase shifts at $m_{\pi} \sim 390$MeV. In \Sec{extrap}, we explain our method for removing the effective range contributions to the data at $m_{\pi} \sim 230$ MeV and from \cite{Torok:2009dg}, and perform an extrapolation of the scattering lengths to the physical quark masses using Heavy Baryon $\chi$PT (HB$\chi$PT), including a discussion about the convergence of the chiral expansion. Finally, we provide a summary of our results and additional conclusions in \Sec{conclusions}.

\begin{table}
\centering
\caption{\label{tab:systems}Meson-baryon systems studied in this work, including isospin (I) and strangeness (S) content.}
\begin{tabular}{|c|c|c|c|}
\hline
System & Quark Content & I & S\\
\hline
$\Xi^0 (\pi^{+})$ & uss(u$\bar{\mbox{d}})$ & $\frac{3}{2}$ & -2\\
$\Sigma^{+} (\pi^{+})$ & uus(u$\bar{\mbox{d}})$ & $1$ & -1\\
$p (K^{+})$ & uud(u$\bar{\mbox{s}})$ & $1$& $1$\\
$n (K^{+})$ & udd(u$\bar{\mbox{s}})$ &$0$ & $1$\\
\hline
\end{tabular}
\end{table}

\section{\label{sec:lattice}Lattice details and analysis}
\subsection{Gauge field configurations and quark propagators}
For this calculation we have used gauge configurations generated by the Hadron Spectrum Collaboration (for details, see Ref.~\cite{Lin:2008pr}) at a spatial lattice spacing of $b_s=0.1227(8)$ fm \cite{Lin:2008pr}. The gauge fields were created using a $n_f=2+1$-flavor anisotropic tadpole-improved clover fermion action \cite{Chen:2000ej} with a Symanzik-improved gauge action \cite{Symanzik:1983dc,Symanzik:1983gh,Luscher:1984xn,Luscher:1985zq}. We use ensembles at two values of the quark masses corresponding to pion masses of $m_{\pi} \sim 390$ MeV and $m_{\pi} \sim 230$ MeV, with a single strange quark mass corresponding to kaon masses of $m_K \sim 543$ MeV and $m_K \sim 465$, respectively. The renormalized anisotropy parameter, $\xi=b_s/b_t = 3.469(11)$, was determined in Ref.~\cite{Beane:2011sc}. For $m_{\pi} \sim 390$ MeV we use several volumes corresponding to ($L=16,20,24,32$), while for $m_{\pi} \sim 230$ MeV we have a single volume, $L=32$. The largest volume for both pion masses has a large temporal extent, $T=256$, while the smaller volumes have $T=128$. To aid in the determination of the ground states we have large ensemble sizes, ranging from about $800 - 2200$ configurations with $\sim 150$ measurements on each configuration. We use the quark propagators from Ref.~\cite{Beane:2009ky,Beane:2009gs,Beane:2009py,Beane:2011pc}, which were generated using the same fermion action as was used for gauge field generation (for more details, see Ref.~\cite{Beane:2009ky}). 

\subsection{\label{sec:analysis}Analysis}

For each configuration, measurements from approximately 150 source locations are used that are averaged over to improve statistics. We also repeat the calculation on each configuration using the opposite parity source and sink interpolators, and include these into the ensemble after performing a time reversal operation. To determine the energy splitting which arises from interactions, $\delta E_0^{(MB)} = E_0^{(MB)} - E_0^{(B)} - E_0^{(M)}$, where $E_0^{(MB)}$, $E_0^{(B)}$, and $E_0^{(M)}$ are the ground-state energies of the meson-baryon, baryon, and meson systems, respectively, we form the following ratio of correlators,
\beq
R_{MB}(t)=\frac{C_{MB}(t)}{C_M(t) C_B(t)} \underset{t\to\infty}{\longrightarrow} \mathcal A e^{-\delta E_0^{MB}t} \ .
\eeq 
The correlators, $C_{MB}$, $C_B$, and $C_M$ are standard two point functions for the meson-baryon, baryon and meson states respectively, and are resampled using the bootstrap method and then used to form an effective mass difference from the following ratio, 
\beq
\mathcal M_{MB}(t)=\ln \left(\frac{R_{MB}(t)}{R_{MB}(t+1)}\right)\ ,
\eeq
which approaches $\delta E_0^{(MB)}$ asymptotically in time. The bootstrap ensemble is used to determine statistical errors on this quantity.

\begin{figure}
\includegraphics[width=0.48\linewidth]{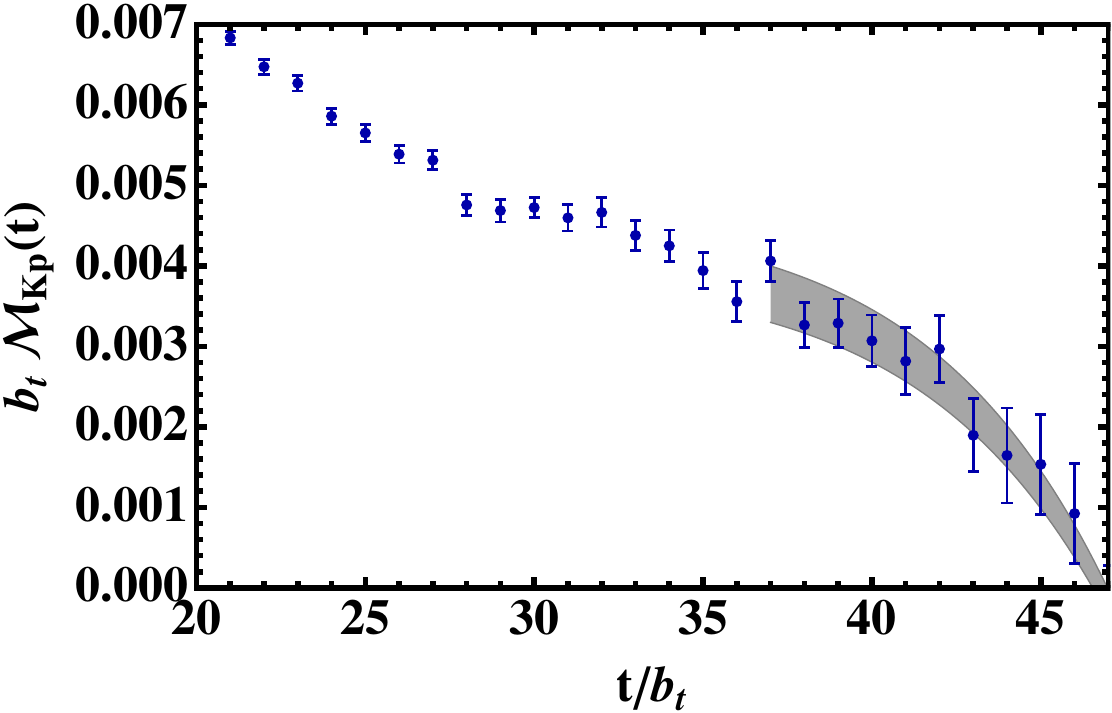}\hspace{1mm}
\includegraphics[width=0.48\linewidth]{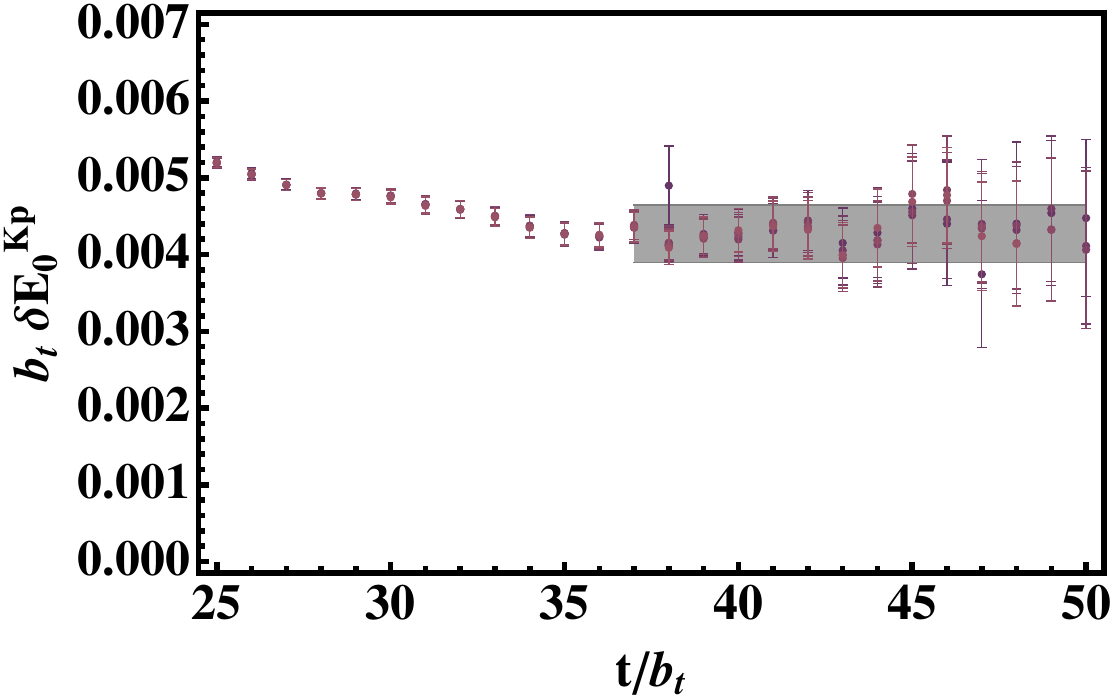} 
\caption{\label{fig:thermal}(Left) Effective mass plot for the kaon-proton system for the $L=20$ ensemble. The gray band is the best fit to a single thermal state plus a single excited state (\Eq{therm}). (Right) Fit results for the ground state energy difference from \Eq{therm} as a function of the time corresponding to the beginning of the fit range, with different colors representing different fit ranges. The band shows the final result for the energy difference, including the uncertainty from the choice of fitting range.}
\end{figure}

The rather short time extent of the three smallest volumes leads to contamination of the correlation functions from thermal effects at times for which the excited state contaminations are still large. In order to extract ground state energies, we have utilized two different types of analysis methods. For the first, we have fit the resampled data directly to a fit function which includes corrections for both the first excited state and the lowest energy thermal state. The leading contribution from thermal effects results from the baryon propagating forward in time and the meson propagating backward. Correspondingly, we have chosen a fit function of the form,
\beq
f_{MB}(t) &=& \ln \left(\frac{g_{MB}(t)}{g_{MB}(t+1)}\right) \ ,\cr
g_{MB}(t) &=& Ae^{-\delta E_0^{MB}t} + B e^{-\delta E_1^{MB} t} + C e^{-m_M (T-2t)}  \ ,
\label{eq:therm}
\eeq
where $\delta E_1^{MB}$ corresponds to the first excited state energy splitting of the meson-baryon system\footnote{Note that the leading effect of including thermal contributions to the pion correlator in the denominator of $R_{MB}(t)$ would be to multiply the entire function $g_{MB}(t)$ by $(1+e^{-m_{\pi}(T-2t)})$, and is therefore exponentially suppressed overall. The fact that we find distinct plateau regions without including this effect is evidence that we may neglect it.}. We use the meson masses extracted from each bootstrap ensemble as input to the correlated fit, and fit the remaining parameters, $\delta E_0^{MB}, \delta E_1^{MB}$, as well as two coefficients (one of the coefficients can be eliminated after inserting $g_{MB}$ into $f_{MB}$), for a chosen time range. This process is repeated over a large set of time ranges, and the extracted values for $\delta E_0^{(MB)}$ are plotted to determine a plateau region. The spread of the fitted $\delta E_0^{MB}$ within $\Delta t = \pm2$ of the plateau region is reported as the fitting systematic error. An example of this procedure is shown in \Fig{thermal}.

\begin{figure}
\includegraphics[width=0.48\linewidth]{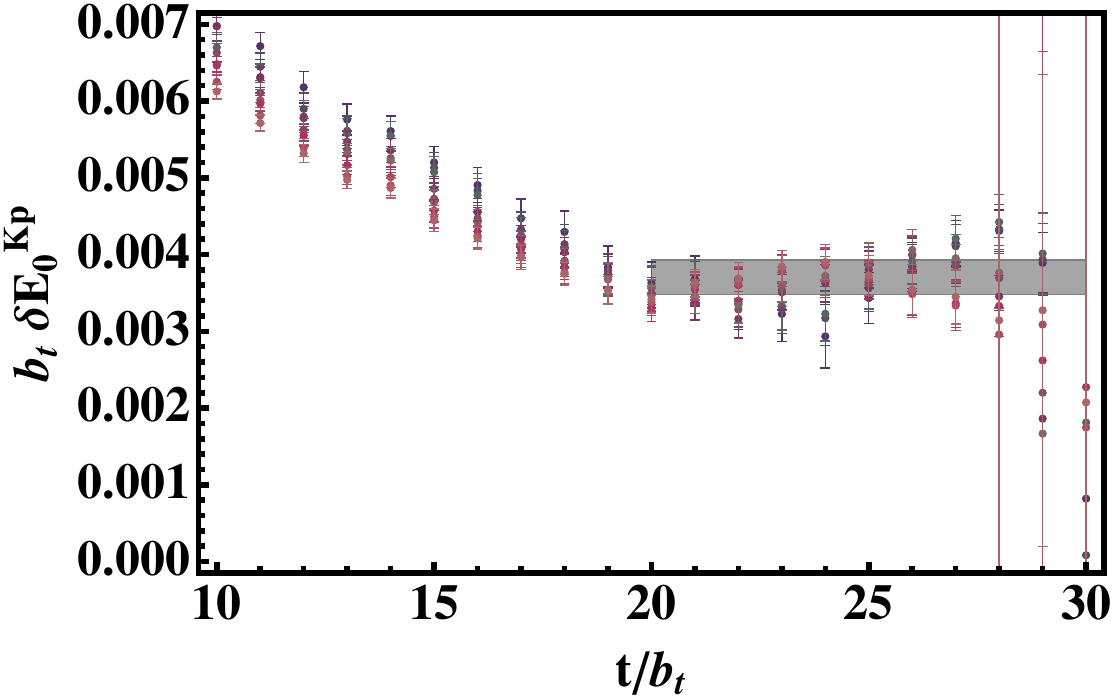} \hspace{1mm}
\includegraphics[width=0.48\linewidth]{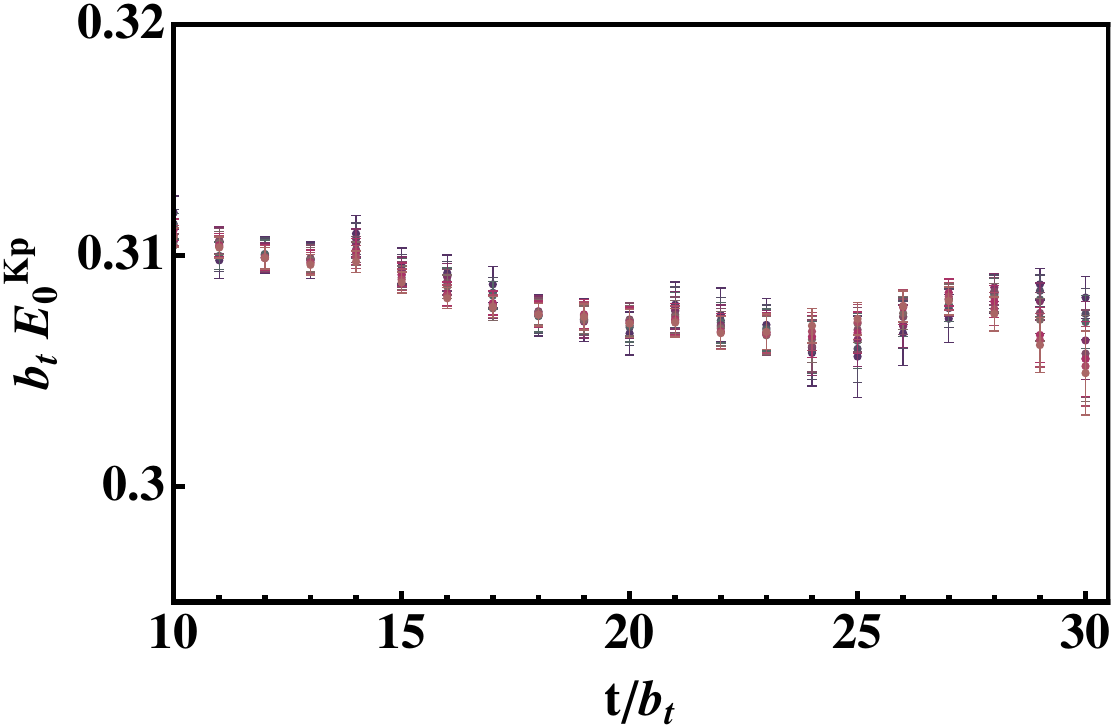}
\includegraphics[width=0.48\linewidth]{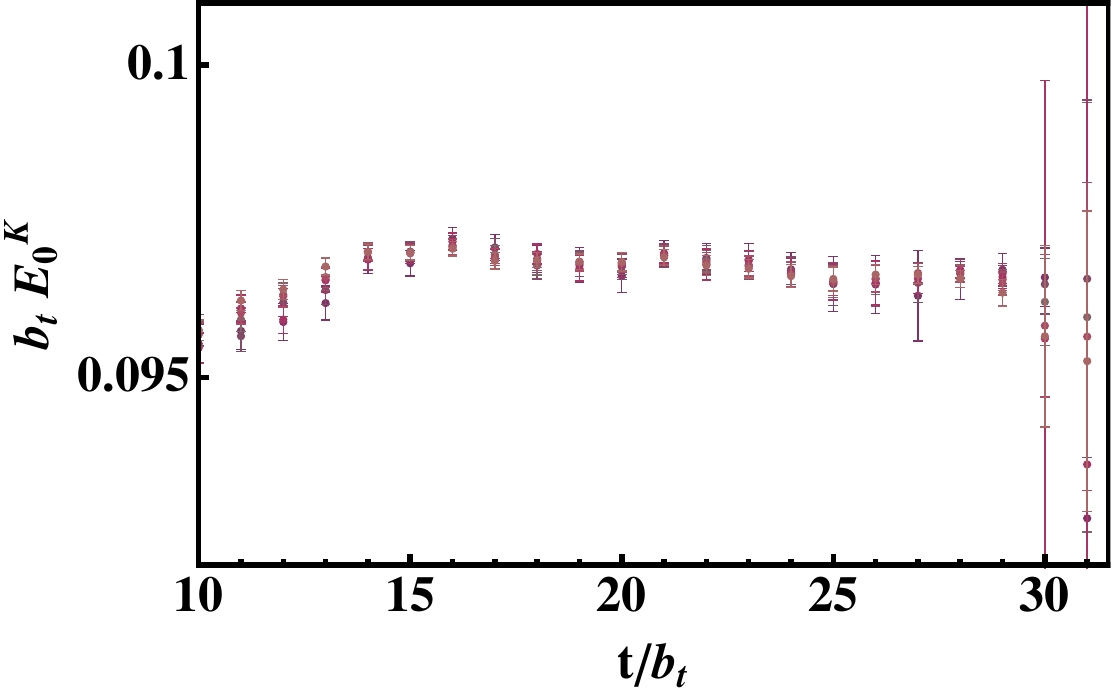}\hspace{1mm}
\includegraphics[width=0.48\linewidth]{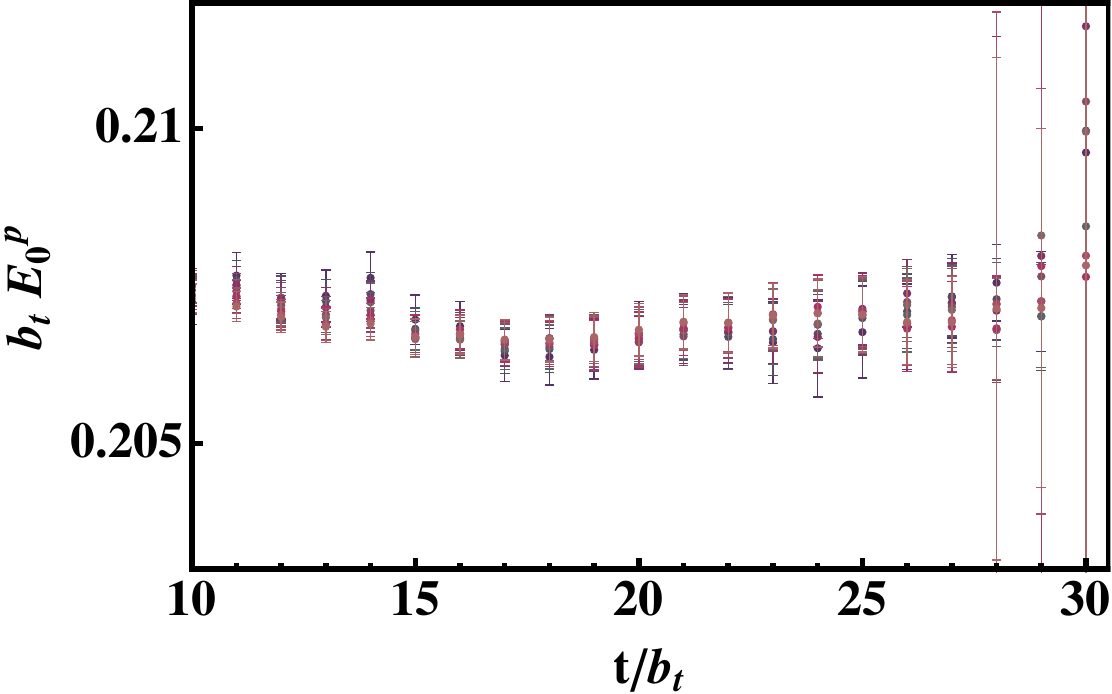}
\caption{\label{fig:Prony}Eigenvalues, $\lambda$, given by the Prony method (\Eq{prony}) for the kaon-proton system for the $L=20$ ensemble as a function of time. Different colors represent different choices of $t_W,t_J$. Clockwise from upper left: Energy difference, kaon-proton energy, proton energy, kaon energy. The band shows a fit result for the energy difference to the plateau at late times.}
\end{figure}

The second analysis method used was the matrix-Prony method \cite{Beane:2009ky}. For this, we use two types of sink for the quark propagators, local and gauge-invariantly smeared, while for both cases we use smeared quark sources. This leads to two sets of correlation functions, labeled smeared-point (SP) and smeared-smeared (SS). The matrix-Prony method uses information from multiple time slices and multiple sources to eliminate excited state contributions to the correlation function. The method solves the following generalized eigenvalue problem,
\beq
M q = \lambda V q \ ,
\label{eq:prony}
\eeq
where $M,V$ are $2\times 2$ matrices formed using the 2-component vector $y$ of correlation functions corresponding to (SS) and (SP), and the eigenvalues $\lambda$ determine the energies. A simple solution exists of the form,
\beq
M = \left[\sum_{\tau=t}^{t+t_W}y(\tau+t_J) y(\tau)^T\right]^{-1}  \hspace{2mm} ,  V = \left[\sum_{\tau=t}^{t+t_W}y(\tau) y(\tau)^T\right]^{-1} \ ,
\eeq
where $t_J$ and $t_W$ may be varied to provide numerical stability, but must be at least 2 for the matrices to achieve full rank. The eigenvalues, $\lambda$, for each of the individual meson, baryon, and meson-baryon correlators are determined as a function of time on each bootstrap sample, and the lowest eigenvalues are combined to produce an energy shift, then plotted to determine a plateau region (see \Fig{Prony}). Fitting systematic uncertainties are obtained by varying this region by $\Delta t=\pm 2$, while statistical uncertainties are determined using the bootstrap ensemble. The parameters $t_J$ and $t_W$ are also varied within $4\leq t_J,t_W \leq 10$, and the spread of the resulting energies is used as a second fitting systematic error, with the total fitting systematic given by the two errors added in quadrature. The various sets of data points (overlapping) in \Fig{Prony} correspond to the different choices of $t_J$ and $t_W$. While the Prony method in principle could also be used to account for thermal effects, as implemented this method leads to ground-state saturation much earlier in time where thermal effects are not significant, as seen in the Figure.

In most cases the two analysis methods gives results for the energies which are compatible within error bars. However, in some cases the central values resulting from the fits to \Eq{therm} are systematically lower than those obtained using the Prony method. For this reason, we report the results for the energies using each fitting method in \Tab{energies}, and perform simultaneous fits to both sets of energies to determine the scattering lengths and effective range parameters, as described in the next Section. We find reasonable agreement for $L=20$ with previous results from Ref.~\cite{Torok:2009dg} using a different action and slightly different quark mass. This suggests that discretization effects are not large.

\section{\label{sec:phaseshifts}Phase shifts at $m_{\pi} \sim 390$ MeV}

We follow L\"{u}scher's method for determining scattering phase shifts using the ground state energies of two particles in a periodic box \cite{Luscher:1986pf,Luscher:1990ux,Briceno:2015tza,Briceno:2014oea,Bernard:2008ax}. For $s$-wave scattering, we use the relation \cite{Beane:2003da}
\beq
p \cot \delta(p) = \frac{1}{\pi L} S\left(\left(\frac{p L}{2\pi}\right)^2\right) \ ,
\eeq
where $\delta(p)$ is the elastic scattering phase shift, and
\beq
S(\eta) = \underset{\Lambda \to \infty}{\mathrm{lim}} \left[ \sum_{\mathbf{j}}^{|\mathbf{j}|< \Lambda} \frac{1}{|\mathbf{j}|^2-\eta^2}-4\pi \Lambda \right] \ .
\eeq
The scattering momenta $p$ of the zero total momentum systems, which are the solution to $\delta E_0^{(MB)} = \sqrt{p^2 + m_M^2}+\sqrt{p^2+m_B^2}-m_M-m_B$, are used as input\footnote{For calculations performed on anisotropic lattices, the energy-momentum relation is modified, so that $\delta E_0^{(MB)} = \sqrt{p^2/\xi_M^2 + m_M^2}+\sqrt{p^2/\xi_B^2+m_B^2}-m_M-m_B$, where $\xi_{M,B}$ is the anisotropy factor for the associated meson (baryon), and energies and masses are given in temporal lattice units, while momenta are given in spatial lattice units. In this work, we use $\xi_M = \xi_B = \xi$, where $\xi$ has been determined using the pion dispersion relation. We find no significant differences in the results using the anisotropy associated with the baryons.}. We fit the resulting phase shifts as a function of $p$ to an effective range expansion (ERE) cut off at NLO,
\beq
p \cot \delta(p) \approx \frac{1}{a} + \frac{1}{2} r_0 p^2 \ .
\eeq
All scattering momenta in these calculations are below the $t$-channel cuts of the respective channels, the lowest of which is at $p^2=m_{\pi}^2/4$, and so the effective range expansion should be a valid representation.

\begin{table}
\centering
\caption{\label{tab:energies}Masses, energy shifts, and phase shifts calculated on each of the four volumes. All quantities are given in lattice units, and errors correspond to statistical and systematic, respectively. For quantities containing two rows of values, the upper row corresponds to the result using a fit function including a thermal plus one excited state (\Eq{therm}), and the lower row corresponds to the result from a fit using the Prony method (\Eq{prony}).}
\begin{tabular}{|c|c|c|c|c|c|}
\hline
&$L=16$&$L=20$&$L=24$&$L=32$&$L=32 \,(230\mathrm{MeV})$\\
\hline
$m_{\pi}$&0.06946(15)(5)&0.06916(11)(2)&0.069077(65)(8)&0.069051(13)(17)&0.039026(80)(12)\\
$m_{K}$&0.097211(31)(33)&0.097015(22)(10)&0.096951(15)(40)&0.096926(9)(13)&0.083092(34)(5)\\
$m_{p}$&0.21029(55)(18)&0.20765(32)(30)&0.20511(18)(31)&0.20473(22)(13)&0.16888(48)(43)\\
$m_{\Sigma}$&0.23016(24)(62)&0.22792(22)(15)&0.22775(18)(5)&0.22796(14)(96)&0.20251(25)(25)\\
$m_{\Xi}$&0.24372(22)(78)&0.24115(36)(2)&0.24042(14)(9)&0.23980(18)(21)&0.22017(29)(32)\\
\hline
\hline
\multirow{2}{*}{$\delta E_0^{(\pi\Sigma)}$}&0.00943(33)(6)& 0.00570(34)(7)& 0.00264(4)(30)& 0.00201(12)(11)&0.00228(24)(8)\\
&0.00836(27)(21)& 0.00469(16)(19)& 0.00232(15)(46)& 0.00196(40)(25)&0.00233(30)(49)\\
\hline
\multirow{2}{*}{$\delta E_0^{(Kp)}$}&0.00682(6)(12)& 0.00427(25)(28)& 0.00274(6)(70)& 0.00241(12)(10)&0.00265(46)(10)\\
&0.00636(40)(20)& 0.00385(19)(12)& 0.00230(68)(72)& 0.00226(19)(22)&0.00233(74)(52)\\
\hline
\multirow{2}{*}{$\delta E_0^{(\pi\Xi)}$}&0.00252(14)(19)& 0.00132(27)(7)& 0.00084(13)(3)& 0.00058(11)(7)&0.00158(18)(1)\\
&0.00222(31)(37)& 0.00126(13)(16)& 0.00061(9)(25)& 0.00065(14)(21)&0.00163(29)(52)\\
\hline
\multirow{2}{*}{$\delta E_0^{(Kn)}$}&0.00192(14)(7)&0.00130(15)(12)&0.00097(6)(17)&0.000717(90)(28)&0.00144(27)(1)\\
&0.00157(28)(52)&0.00112(6)(16)&0.00074(11)(26)&0.00069(16)(12)&0.00112(60)(34)\\
\hline
\hline
\multirow{2}{*}{$p \cot \delta^{(\pi\Sigma)}$}&-0.3883(56)(83)&-0.3267(83)(53)&-0.358(27)(8)&-0.2256(94)(89)&-0.286(21)(7)\\
&-0.423(38)(10)&-0.376(42)(6)&-0.41(11)(2)&-0.233(57)(11)&-0.299(21)(55)\\
\hline
\multirow{2}{*}{$p \cot \delta^{(Kp)}$}&-0.426(12)(17)&-0.348(42)(3)&-0.307(16)(2)&-0.1690(51)(58)&-0.178(19)(10)\\
&-0.448(24)(6)&-0.374(17)(3)&-0.35(13)(2)&-0.177(18)(7)&-0.194(25)(29)\\
\hline
\multirow{2}{*}{$p \cot \delta^{(\pi\Xi)}$}&-1.082(71)(88)&-1.03(39)(7)&-0.93(17)(4)&-0.58(11)(3)&-0.377(35)(2)\\
&-1.20(16)(16)&-1.07(22)(3)&-1.30(50)(41)&-0.53(22)(2)&-0.37(10)(11)\\
\hline
\multirow{2}{*}{$p \cot \delta^{(Kn)}$}& -1.16(88)(11)&-0.883(96)(11)&-0.690(48)(60)&-0.414(58)(18)&-0.326(50)(34)\\
&-1.37(29)(26)&-0.99(14)(2)&-0.87(25)(23)&-0.43(17)(2)&-0.333(69)(72)\\
\hline
\end{tabular}
\end{table}

Using the two methods outlined in \Sec{analysis}, we extract two values for the ground state energy for a given system at each volume. Bootstrap ensembles of the energies are used to calculate the corresponding $p^2$, which we label $\eta_{L,f,i}$ and $p\cot\delta$, which we label $\delta_{L,f,i}$, corresponding to bootstrap ensemble $i$, volume $L$, and fitting method $f$. In certain cases one analysis method is clearly more successful than the other, as evidenced by the associated q-value of the fit and large systematic errors. For this reason, a fit is performed to all of the data, including correlation between the two analysis methods, by minimizing the following correlated, weighted $\chi^2$,
\beq
\chi_i^2 \propto \sum_{\{f,g\}=1}^2 \sum_{L=1}^4 \sqrt{Q_{L,f} Q_{L,g}} \left(\delta_{L,f,i}-(1/a+\frac{1}{2} r_0 \eta_{L,f,i})\right)\left[ \mathcal C_{(L)}^{-1}\right]_{f,g} \left( \delta_{L,g,i}-(-1/a+\frac{1}{2} r_0 \eta_{L,g,i})\right) \ ,
\eeq
where $\left[ \mathcal C_{(L)} \right]_{fg}= \langle \left( \delta_{L,f} - \langle \delta_{L,f} \rangle \right)\left( \delta_{L,g} - \langle \delta_{L,g} \rangle \right) \rangle$, with angle brackets denoting an average over the bootstrap ensemble, is the covariance matrix encoding correlations between different analysis methods for the same volume, $L$, and $Q_{L,f}$ is the q-value for fit $f$ on volume $L$ and acts as a weight. Each ERE fit is performed using the energy results from the thermal fit and a given Prony fit. An ensemble of such ERE fits using different Prony results and different time windows is used to calculate the systematic uncertainty. 

\begin{figure}
\includegraphics[width=0.48\linewidth]{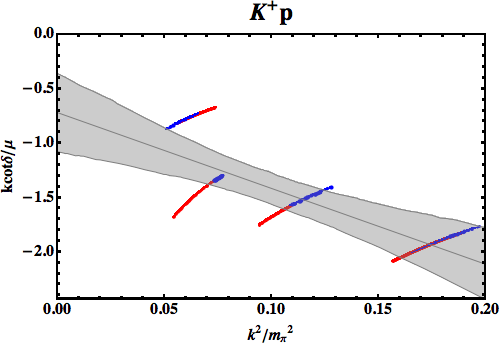}\hspace{1mm}
\includegraphics[width=0.48\linewidth]{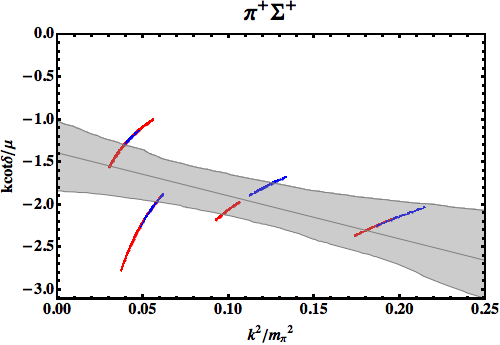} \\

\vspace{2mm}

\includegraphics[width=0.48\linewidth]{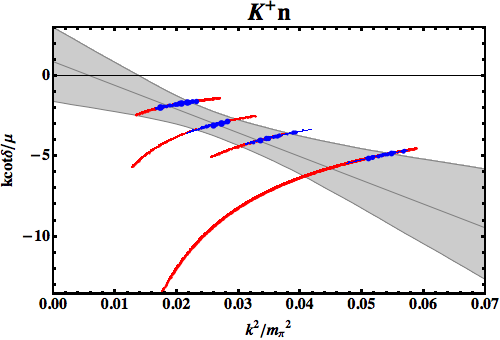} \hspace{1mm}
\includegraphics[width=0.48\linewidth]{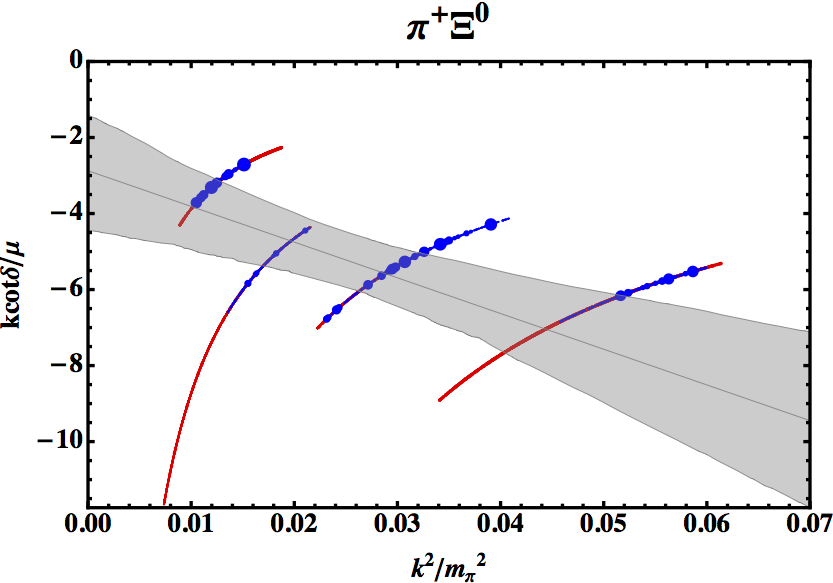}
\caption{\label{fig:phaseshifts}Phase shifts for meson-baryon systems as a function of the scattering momentum. Red points (appearing as curves due to their large overlap) were extracted on each bootstrap sample using the Prony method (\Eq{prony}) for various $t_J,t_W$, while blue points resulted from thermal plus excited state fits (\Eq{therm}) on each bootstrap sample. The relative sizes of the points denote the quality of the fits used to generate them, and therefore, their relative contributions to the fit to an effective range expansion, cut off at $\mathcal O(k^2)$ (gray band).}
\end{figure}

\begin{figure}
\includegraphics[width=0.48\linewidth]{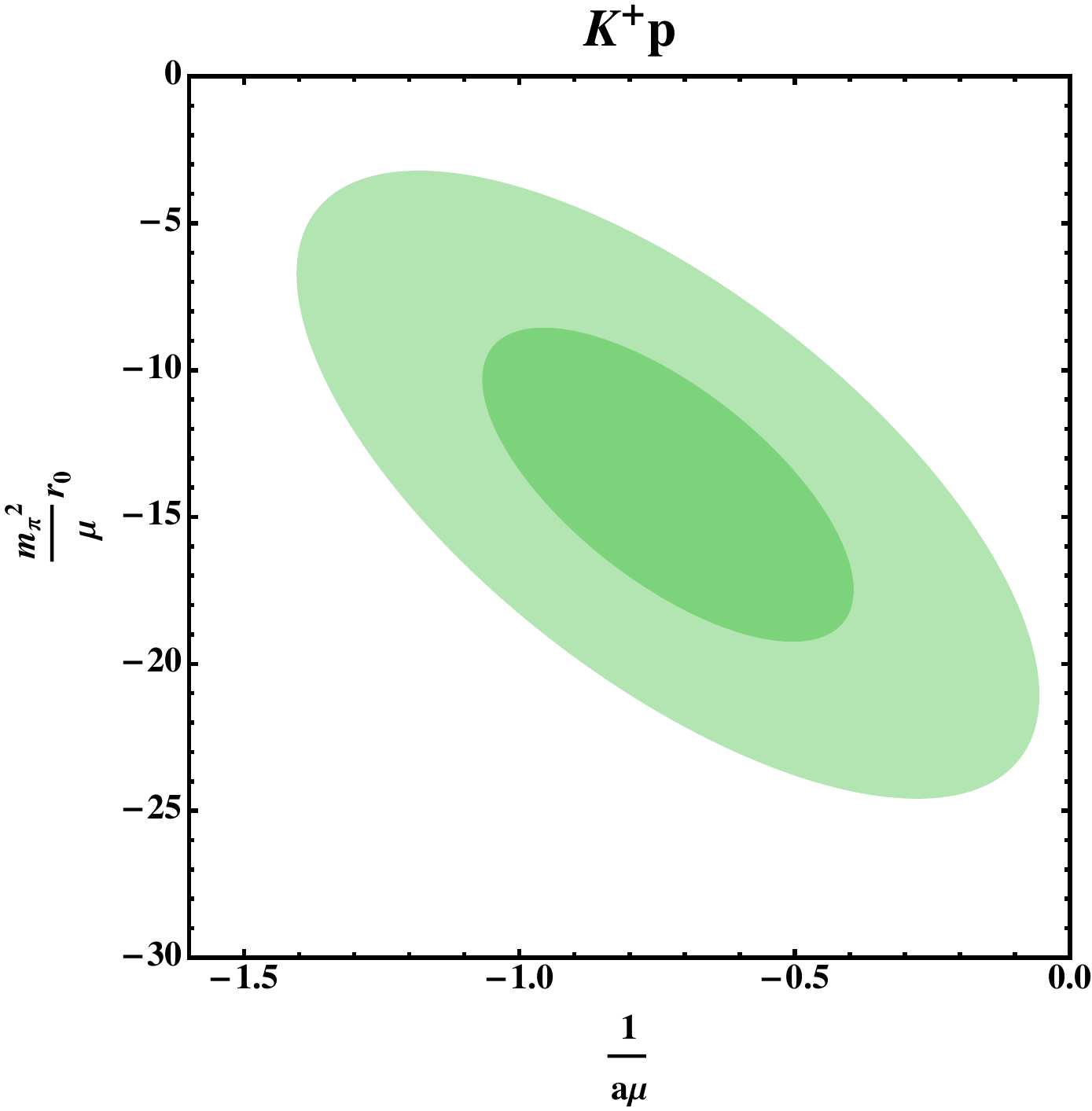}\hspace{1mm}
\includegraphics[width=0.48\linewidth]{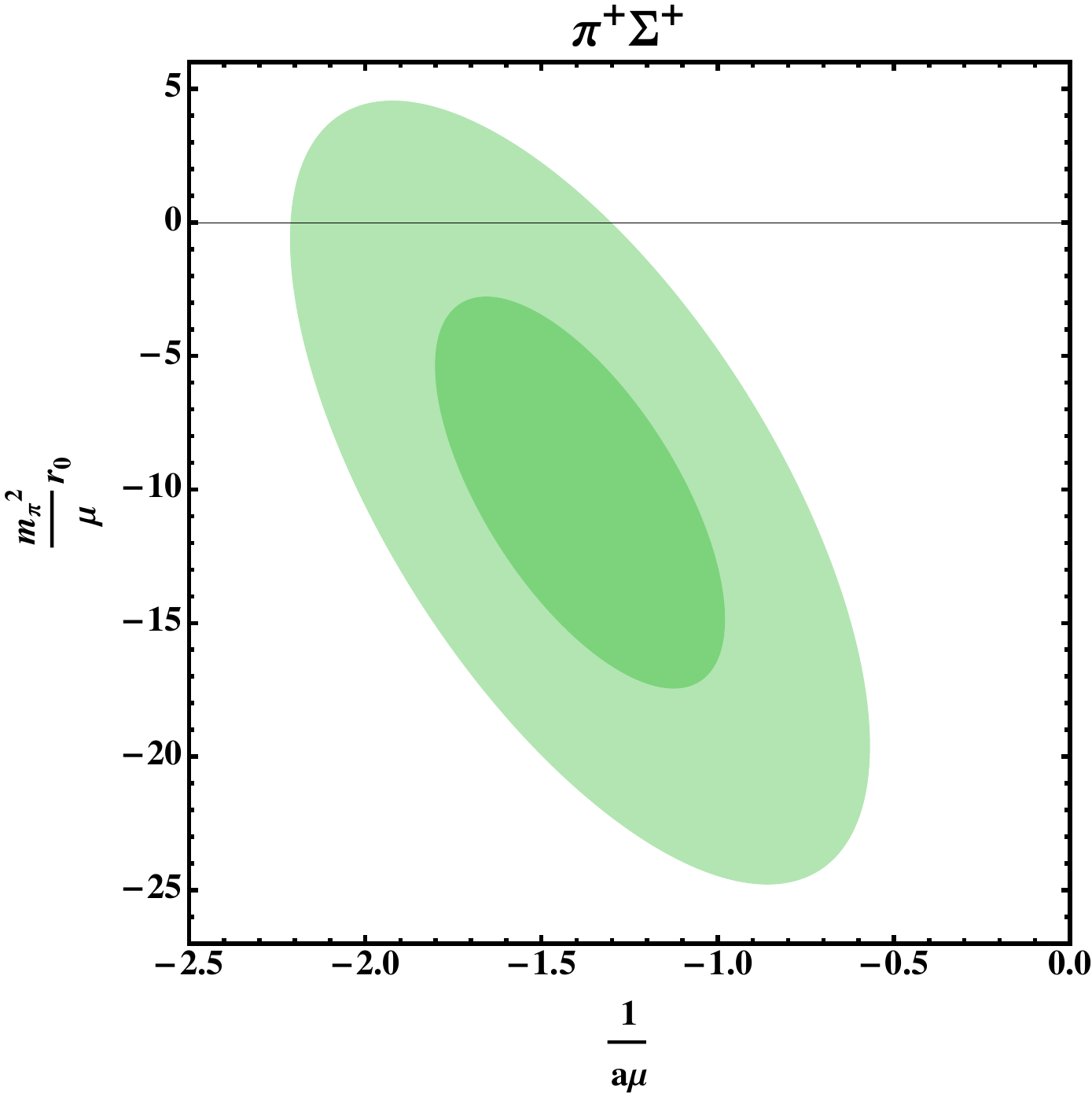} \\

\vspace{2mm}

\includegraphics[width=0.48\linewidth]{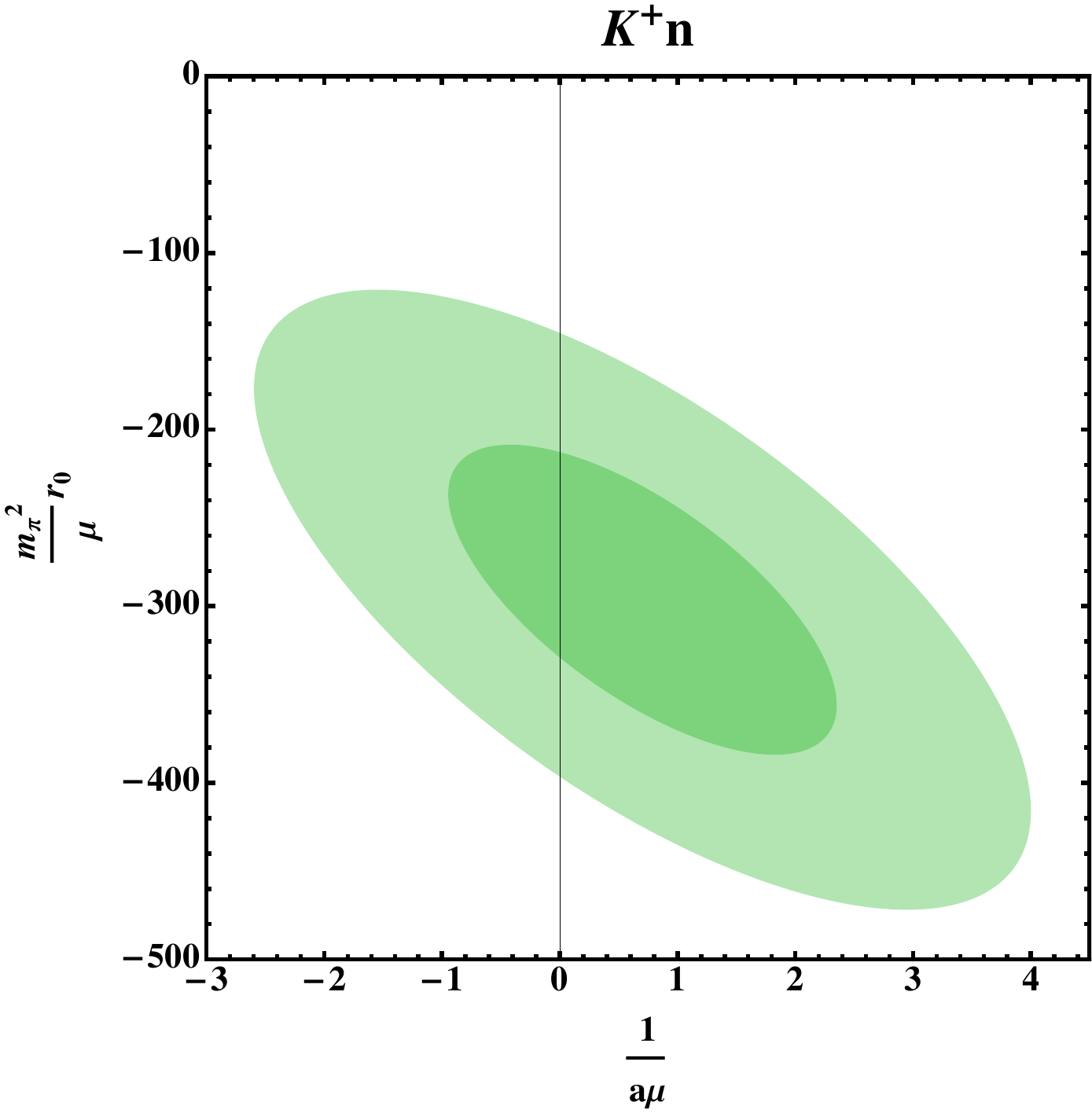} \hspace{1mm}
\includegraphics[width=0.48\linewidth]{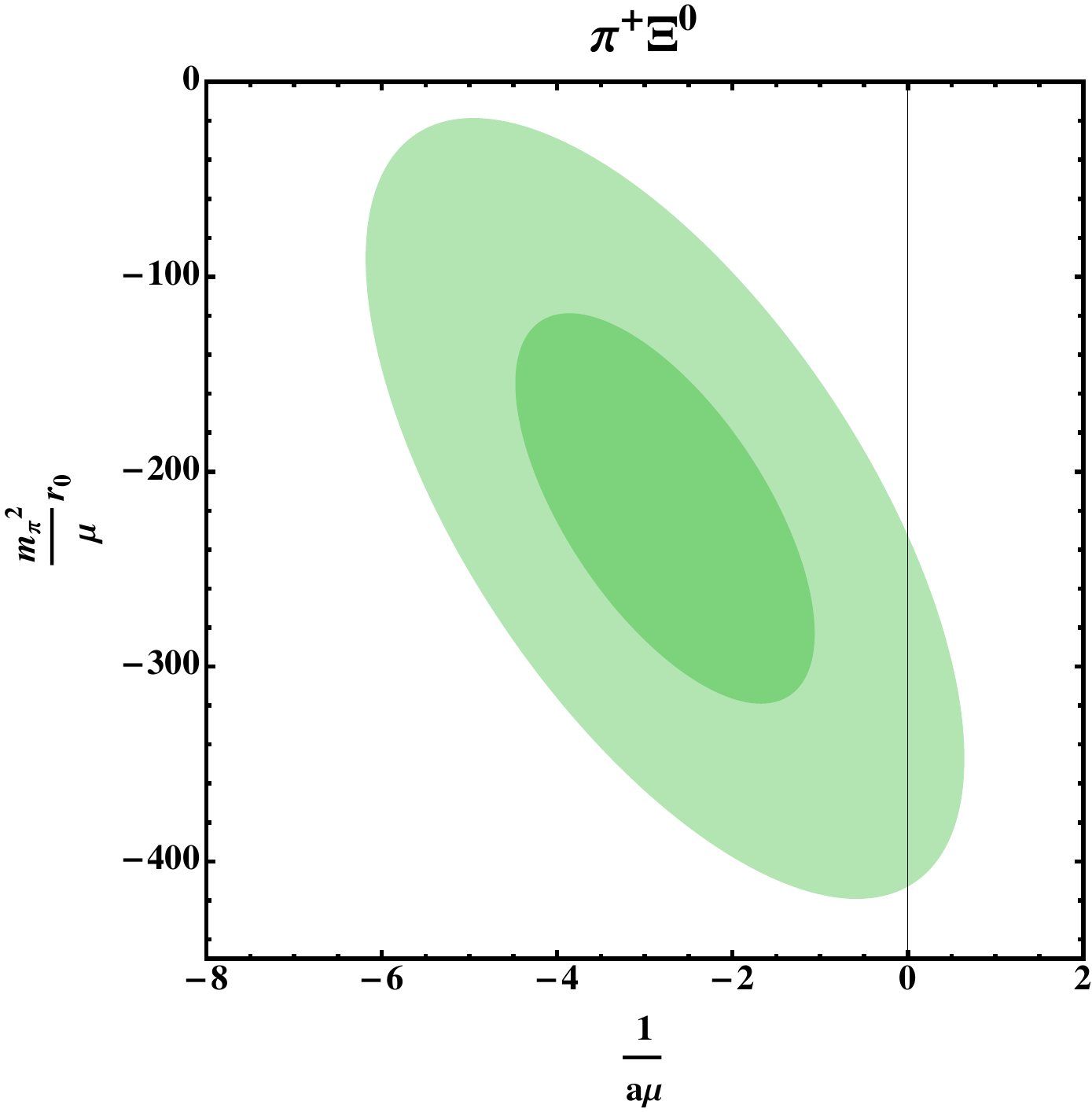}
\caption[]{\label{fig:errellipses}Error ellipses for the scattering lengths and effective ranges of the meson-baryon systems in units corresponding to the intercepts and slopes of \Fig{phaseshifts}. The inner and outer bands represent $68\%$ and $95\%$ confidences, respectively.}
\end{figure}

Results for the fits to the effective range expansion are shown in \Fig{phaseshifts}, with numerical values given in \Tab{ERE}. The bands that are shown include both statistical and fitting systematic uncertainties, added in quadrature. In \Fig{errellipses}, we plot error ellipses for the extracted scattering lengths and effective ranges. Note that the effective ranges are given in units of $m_{\pi}$. Thus we find that the effective ranges are significantly larger than the naive expectation, $r_0 \sim 1/m_{\pi}$, particularly for $K^{+}n$ and $\pi^{+}\Xi^0$. Therefore, the effective range contributions, especially for the smaller volumes that correspond to the largest scattering momenta, should not be neglected for an accurate determination of the scattering lengths. The large values of the effective range parameters may also call into question the validity of the L\"uscher analysis, as we will discuss below. Note that the $\pi^{+}\Xi^0$ ($\pi^{+}\Sigma^{+}$) channel is related to the $K^{+}n$ ($K^{+}p$) channel by isospin. This near-symmetry is reflected in the scattering lengths and effective ranges, thus, once an anomalously large effective range is found in the $\pi^{+}\Xi^0$, it is not surprising that the $K^{+}n$ channel also displays a large effective range. 

\begin{table}
\centering
\caption{\label{tab:ERE}Fit results for the scattering lengths and effective ranges of the meson-baryon systems, as well as the correlation coefficient between the two quantities. Both statistical and systematic uncertainties are included in the quoted error bar.}
\begin{tabular}{|c|c|c|c|c|}
\hline
& $\pi^{+}\Sigma^{+}$ & $K^{+}p$  & $\pi^{+}\Xi^0$& $K^{+}n$\\
\hline
$\frac{1}{a\mu}$ &-1.39(31) &-0.73(25) &-2.8(13) &0.7(12) \\
$\frac{m_{\pi}^2}{\mu}r_0$ &-10.1(73) &-13.9(54) &-220(100) &-296(88)\\
$\rho_{\frac{1}{a\mu},\frac{m_{\pi}^2}{\mu}r_0} $ & -0.85 & -0.91 &-0.83 & -0.93 \\ 
\hline
\end{tabular}
\end{table}

The large effective ranges that we find may lead to concern that the volumes used are not sufficiently large for this calculation. However, the effective range, like the scattering length, is simply a parameter in an ERE expansion, and needs not be smaller than the box size in order for the L\"uscher method to be valid \cite{Beane:2003da}. Leading corrections to the L\"uscher relation due to the finite range of the interaction scale as $e^{-L/r_\mathrm{int}}$, and are of order $10^{-4} - 10^{-7}$ for the volumes considered in this work. Thus, it is unlikely that the L\"uscher method breaks down for these calculations. 

\begin{figure}
\includegraphics[width=0.48\linewidth]{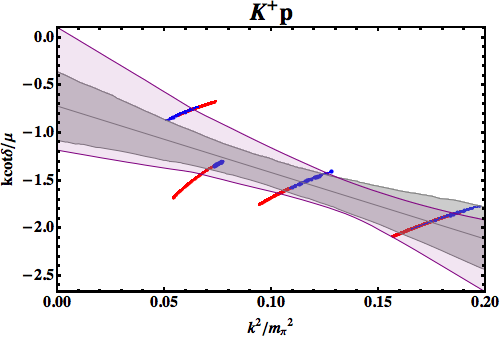}\hspace{1mm}
\includegraphics[width=0.48\linewidth]{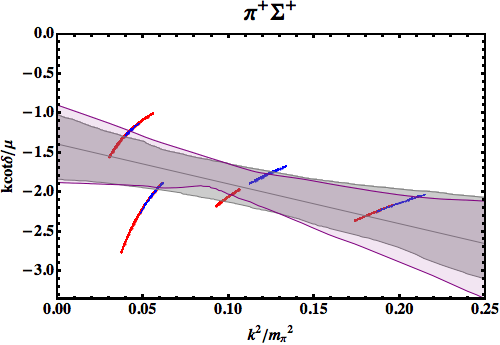} \\

\vspace{2mm}

\includegraphics[width=0.48\linewidth]{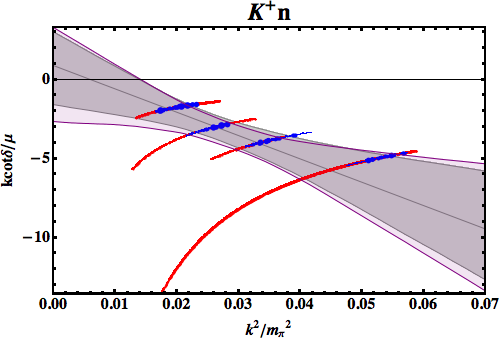} \hspace{1mm}
\includegraphics[width=0.48\linewidth]{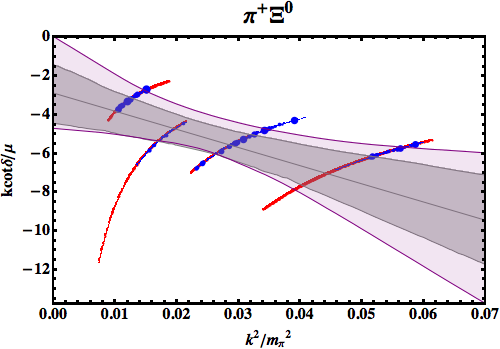}
\caption{\label{fig:phaseshifts_rmL}Phase shifts for meson-baryon systems as a function of the scattering momentum, with data points and gray band as described in \Fig{phaseshifts}. Purple band shows the results from performing a set of fits to an effective range expansion, cut off at $\mathcal O(k^2)$. The set includes all possible fits performed after removing the data corresponding to a single volume, for each volume.}
\end{figure}

An inspection of \Fig{phaseshifts} could also raise a concern that the values extracted for the effective ranges hinge disproportionately on the fit results for a single volume. For example, the smallest volume, corresponding to $L=16$, gives the largest leverage in $k^2$, and would be the most affected by potential contributions from exponential finite volume effects. On the other hand, the largest volume, $L=32$, in some cases appears to provide significant leverage toward large effective ranges, and is the most precise and least subject to exponential volume effects. As an additional test to determine whether the results from any single volume exert a disproportionate influence over the extracted effective ranges, we have performed all possible fits in which the data from a single volume has been removed. The resulting error bands are overlaid in purple in \Fig{phaseshifts_rmL}. While the error bars on the scattering lengths and effective ranges become slightly larger because of the lower number of degrees of freedom for each fit, the results do not change significantly (we do not include this test in our quoted errors bars on the effective range parameters). Therefore, we conclude that the large effective ranges that we find are likely not due to finite volume effects from the smallest volume, or to anomalously large $k^2$ values for the $L=32$ fits.

\section{\label{sec:extrap}Extrapolation to the physical pion mass}

Using SU(3) HB$\chi$PT to next-to-next-to-leading order, the scattering lengths of the pion-baryon systems have been determined in Ref.~\cite{Liu:2006xja,Liu:2007ct} to be,
\beq
a_{\pi^{+}\Sigma^{+}} &=& \frac{1}{4\pi} \frac{m_{\Sigma}}{m_{\pi}+m_{\Sigma}} \left[ -\frac{2m_{\pi}}{f_{\pi}^2}+\frac{2m_{\pi}^2}{f_{\pi}^2}C_1 + \mathcal Y_{\pi^{+} \Sigma^{+}}(\mu)+8h_{123}(\mu)\frac{m_{\pi}^3}{f_{\pi}^2} \right] \ , \cr
a_{\pi^{+}\Xi^{0}} &=& \frac{1}{4\pi} \frac{m_{\Xi}}{m_{\pi}+m_{\Xi}} \left[ -\frac{m_{\pi}}{f_{\pi}^2}+\frac{m_{\pi}^2}{f_{\pi}^2}C_{01} + \mathcal Y_{\pi^{+} \Xi^{0}}(\mu)+8h_{1}(\mu)\frac{m_{\pi}^3}{f_{\pi}^2} \right] \ ,
\eeq
where $C_{01} \equiv C_0 + C_1$ and $h_{123} \equiv h_1-h_2+h_3$, and $C_0, C_1, h_1, h_2,$ and $h_3$ are low-energy constants (LECs) of the HB$\chi$PT. The scattering length formulae for the kaon-baryon systems are produced by making the replacements $\{ a_{\pi^{+}\Sigma^{+}},m_{\pi}, f_{\pi}, m_{\Sigma}, \mathcal Y_{\pi^{+}\Sigma^{+}} \} \leftrightarrow \{ a_{K^{+}p} ,m_{K}, f_{K}, m_{p}, \mathcal Y_{K^{+}p} \} $ and $\{ a_{\pi^{+}\Xi^{+}},m_{\pi}, f_{\pi}, m_{\Xi}, \mathcal Y_{\pi^{+}\Xi^{+}} \} \leftrightarrow \{a_{K^{+}n} ,m_{K}, f_{K}, m_{n}, \mathcal Y_{K^{+}n} \} $. The loop functions are defined by
\beq
\mathcal Y_{\pi^{+}\Sigma^{+}}(\mu) &=& \frac{m_{\pi}^2}{2\pi^2 f_{\pi}^4}\left[-m_{\pi}\left(\frac{3}{2}-2\ln \frac{m_{\pi}}{\mu}-\ln \frac{m_K}{\mu}\right)-\sqrt{m_K^2-m_{\pi}^2}\cos^{-1}\frac{m_{\pi}}{m_K} \right.\cr
&+&\left. \frac{\pi}{2}\left(3F^2m_{\pi}-\frac{1}{3}D^2m_{\eta}\right)\right] \ , \cr
\mathcal Y_{\pi^{+}\Xi^{0}}(\mu) &=& \frac{m_{\pi}^2}{4\pi^2 f_{\pi}^4}\left[-m_{\pi}\left(\frac{3}{2}-2\ln \frac{m_{\pi}}{\mu}-\ln \frac{m_K}{\mu}\right)-\sqrt{m_K^2-m_{\pi}^2}\left(\pi+\cos^{-1}\frac{m_{\pi}}{m_K}\right) \right.  \cr
&+& \left.\frac{\pi}{4}\left(3(D-F)^2m_{\pi}-\frac{1}{3}(D+3F)^2m_{\eta}\right)\right] \ ,\cr
\mathcal Y_{K^{+}p}(\mu) &=& \frac{m_{K}^2}{4\pi^2 f_K^4}\left[m_K\left(-3+2\ln \frac{m_{\pi}}{\mu}+\ln \frac{m_K}{\mu}+3\ln \frac{m_{\eta}}{\mu}\right)+2 \sqrt{m_K^2-m_{\pi}^2}\ln\frac{m_K+\sqrt{m_K^2-m_{\pi}^2}}{m_{\pi}} \right.\cr
&-&\left.3\sqrt{m_{\eta}^2-m_K^2}\cos^{-1}\frac{m_K}{m_{\eta}} -\frac{\pi}{6}(D-3F)\left(2(D+F)\frac{m_{\pi}^2}{m_{\eta}+m_{\pi}}+(D+5F)m_{\eta}\right)\right] \cr
\mathcal Y_{K^{+}n}(\mu) &=& \frac{\mathcal Y_{K^{+}p}}{2}+ \frac{3m_{K}^2}{8\pi^2 f_K^4}\left[m_K\left(\ln \frac{m_{\pi}}{\mu}-\ln \frac{m_K}{\mu}\right)+ \sqrt{m_K^2-m_{\pi}^2}\ln\frac{m_K+\sqrt{m_K^2-m_{\pi}^2}}{m_{\pi}} \right.\cr
&+&\left.\frac{\pi}{3}(D-3F)\left((D+F)\frac{m_{\pi}^2}{m_{\eta}+m_{\pi}}+\frac{1}{6}(7D+3F)m_{\eta}\right)\right] \ .
\eeq
The pion and kaon decay constants are taken from Ref.~\cite{Durr:2013goa,Durr:2010hr}.

Due to the poor convergence for the scattering lengths using $SU(3)$ HB$\chi$PT noted in Ref.~\cite{Torok:2009dg}, we also investigate the $SU(2)$ scattering formulae for the pion-baryon systems,
\beq
a^{SU(2)}_{\pi^{+}\Sigma^{+}} = \frac{1}{4\pi}\frac{m_{\Sigma}}{m_{\pi}+m_{\Sigma}}\left[-\frac{2m_{\pi}}{f_{\pi}^2}+\frac{2m_{\pi}^2}{f_{\pi}^2}C_{\pi^{+}\Sigma^{+}}+\frac{m_{\pi}^3}{\pi^2f_{\pi}^4}\ln\frac{m_{\pi}}{\mu}+\frac{2m_{\pi}^3}{f_{\pi}^2}h_{\pi^{+}\Sigma^{+}}(\mu)\right] \ , \cr
a^{SU(2)}_{\pi^{+}\Xi^{0}} = \frac{1}{4\pi}\frac{m_{\Xi}}{m_{\pi}+m_{\Xi}}\left[-\frac{m_{\pi}}{f_{\pi}^2}+\frac{m_{\pi}^2}{f_{\pi}^2}C_{\pi^{+}\Xi^{0}}+\frac{m_{\pi}^3}{2\pi^2f_{\pi}^4}\ln\frac{m_{\pi}}{\mu}+\frac{m_{\pi}^3}{f_{\pi}^2}h_{\pi^{+}\Xi^{0}}(\mu)\right]
\eeq

Following \cite{Torok:2009dg}, we form the following quantities,
\beq
\Gamma_{NLO}^{\pi\Sigma} &\equiv& -\frac{2\pi a_{\pi\Sigma}f_{\pi}^2}{m_{\pi}}\left(1+\frac{m_{\pi}}{m_{\Sigma}}\right) = 1-C_1m_{\pi} \cr
\Gamma_{NNLO}^{\pi\Sigma} &\equiv& -\frac{2\pi a_{\pi\Sigma}f_{\pi}^2}{m_{\pi}}\left(1+\frac{m_{\pi}}{m_{\Sigma}}\right) +\frac{f_{\pi}}{2m_{\pi}} \mathcal Y_{\pi\Sigma}(\mu)= 1-C_1m_{\pi} -4 h_{123}(\mu)m_{\pi}^2 \cr
\Gamma_{NLO}^{\pi\Xi} &\equiv& -\frac{4\pi a_{\pi\Xi}f_{\pi}^2}{m_{\pi}}\left(1+\frac{m_{\pi}}{m_{\Xi}}\right) = 1-C_{01}m_{\pi} \cr
\Gamma_{NNLO}^{\pi\Xi} &\equiv& -\frac{4\pi a_{\pi\Xi}f_{\pi}^2}{m_{\pi}}\left(1+\frac{m_{\pi}}{m_{\Xi}}\right) +\frac{f_{\pi}}{m_{\pi}} \mathcal Y_{\pi\Xi}(\mu)= 1-C_{01}m_{\pi} -8 h_{1}(\mu)m_{\pi}^2
\label{eq:Gamma}
\eeq
$\Gamma^{Kp}$ and $\Gamma^{Kn}$ may be found by replacing $\{m_{\pi},f_{\pi},m_{\Sigma} ,m_{\Xi},a_{\pi\Sigma},a_{\pi\Xi},\mathcal Y_{\pi\Sigma},\mathcal Y_{\pi\Xi}\} \leftrightarrow \{m_{K},f_{K},m_{p},m_n, a_{Kp},a_{Kn},\mathcal Y_{Kp},\mathcal Y_{Kn}\}$. The $SU(2)$ equivalents of the NNLO quantities, $\Gamma_{\mathrm{NNLO}}^{\mathrm{SU(2)}}$, may be formed in an analogous way.

In addition to the phase shift points calculated in this work, we also include results from Ref.~\cite{Torok:2009dg} in the chiral fits. These were performed at four values of the pion mass at a single volume corresponding approximately to $L=20$ in the current study. Due to the large effective range contributions to the phase shifts that we have found at this volume, we choose to account for this by adding a correction to the results of Ref.~\cite{Torok:2009dg},
\beq
\frac{1}{a(m_{\pi})} = \frac{1}{a_{\mathrm{L}}(m_{\pi})}-\frac{1}{2}r_0(m_{\pi}) p^2 \ , 
\eeq
where $a_{\mathrm{L}}(m_{\pi})$ is the result quoted for the scattering length in Ref.~\cite{Torok:2009dg}. To determine the effective range for a given pion mass, we use the ansatz,
\beq
\label{eq:r0mpi}
r_0(m_{\pi}) = C_r/m_{\pi} \ ,
\eeq
where $C_r$ is a constant to be determined, and which follows from the expected behavior of the effective range in the chiral limit\footnote{Note that though the large effective ranges found for these systems seem to indicate fine-tuning of the potential arising from meson-exchange and contact diagrams, as the pion mass is decreased toward the chiral limit the long-range meson exchange contributions will dominate, leading to the pion mass dependence of \Eq{r0mpi}.}. Because the scattering length and effective range for a given fit are highly correlated, to determine the error on $r_0$ we generate fake data for $\delta E^{(MB)}$ from a Gaussian distribution according to the results reported in Ref.~\cite{Torok:2009dg}. From this, we calculate a (non-Gaussian) distribution for $a_{\mathrm{L}}$. We then form the correlation matrix between $a_{\mathrm{L}}$ and $1/r_0$ using the data from $m_{\pi}\sim 390$ MeV. Because both $a_{\mathrm{L}}$ and $1/r_0$ are proportional to $m_{\pi}$ to leading order in $\chi$PT, the correlation matrix should have minimal dependence on $m_{\pi}$. Finally, using the distribution for $1/a_{\mathrm{L}}$ and the correlation matrix, we generate fake data for $r_0(m_{\pi})$. Using these ensembles we determine the error on $a(m_{\pi})$.

We also perform the same shift for our $m_{\pi}\sim 230$ MeV, $L=32$ data. Because $L=32$ for $m_{\pi}\sim 390$ MeV appears to be a sufficiently large volume for obtaining threshold, and because the data is systematically higher than the fit result at the same volume, we use both the shifted and unshifted $m_{\pi}\sim 230$ MeV data, and include the difference between them in our estimate of the systematic error for these points.

\begin{figure}
\includegraphics[width=0.48\linewidth]{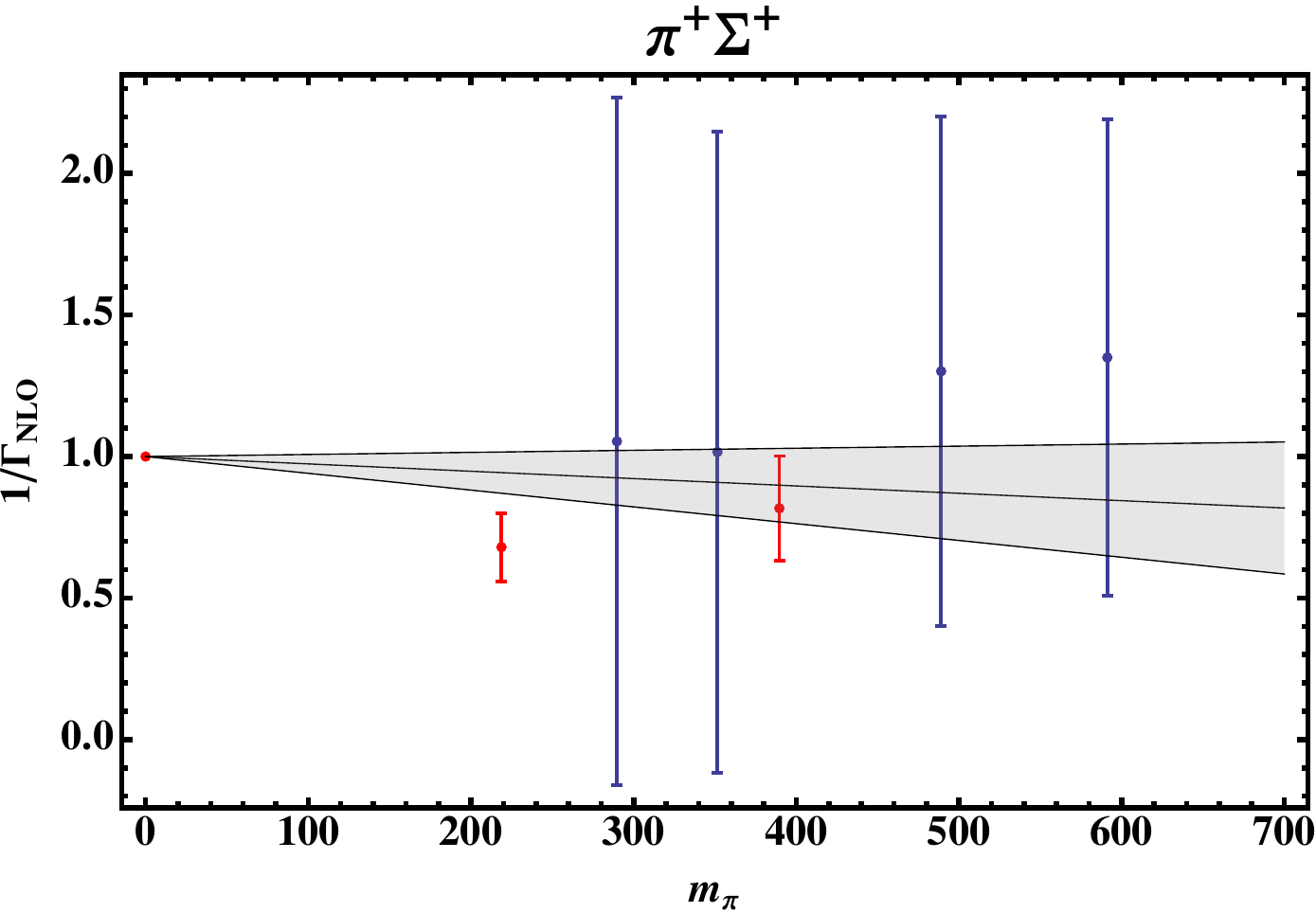}\hspace{1mm}
\includegraphics[width=0.48\linewidth]{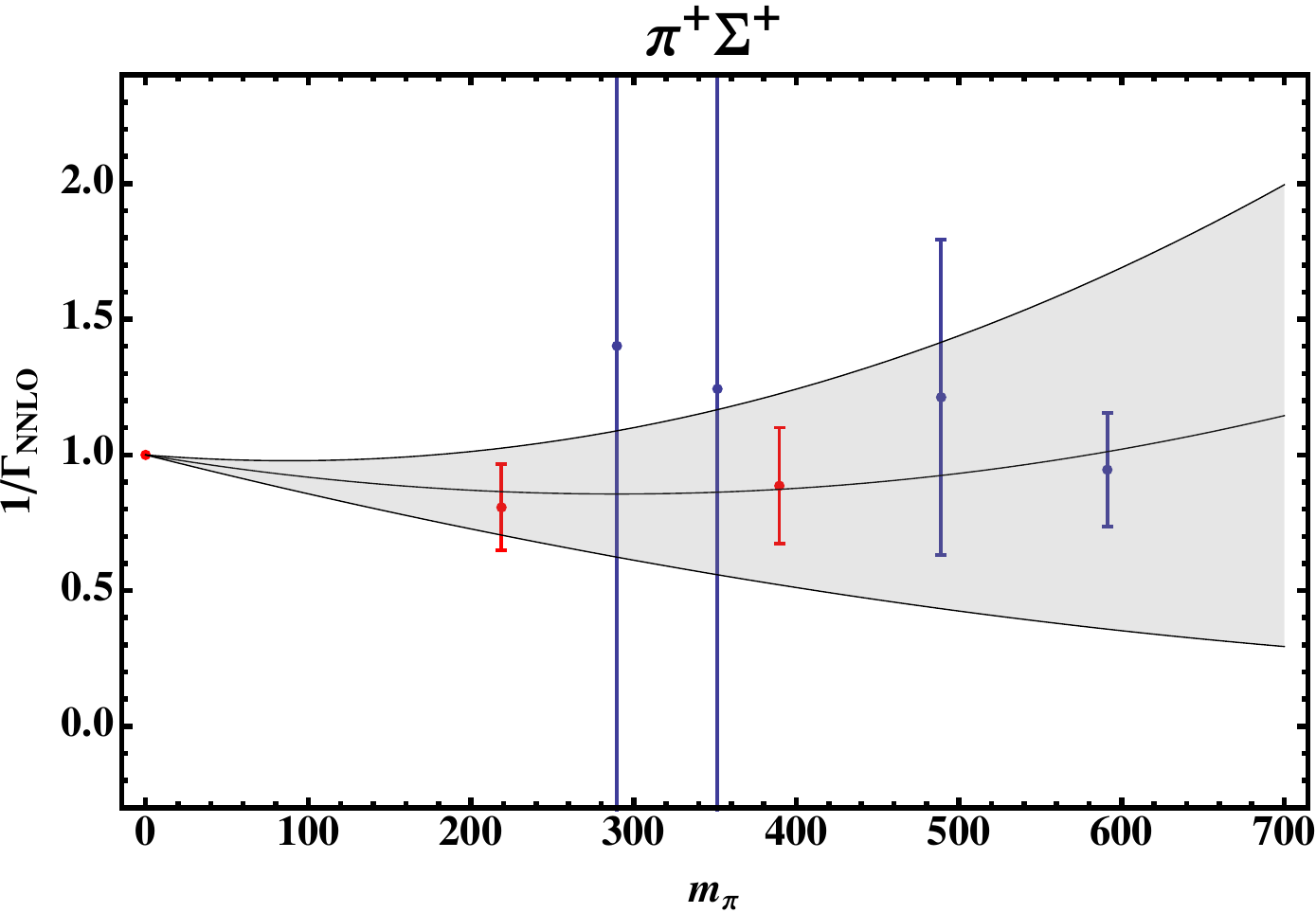} \\

\vspace{2mm}

\includegraphics[width=0.48\linewidth]{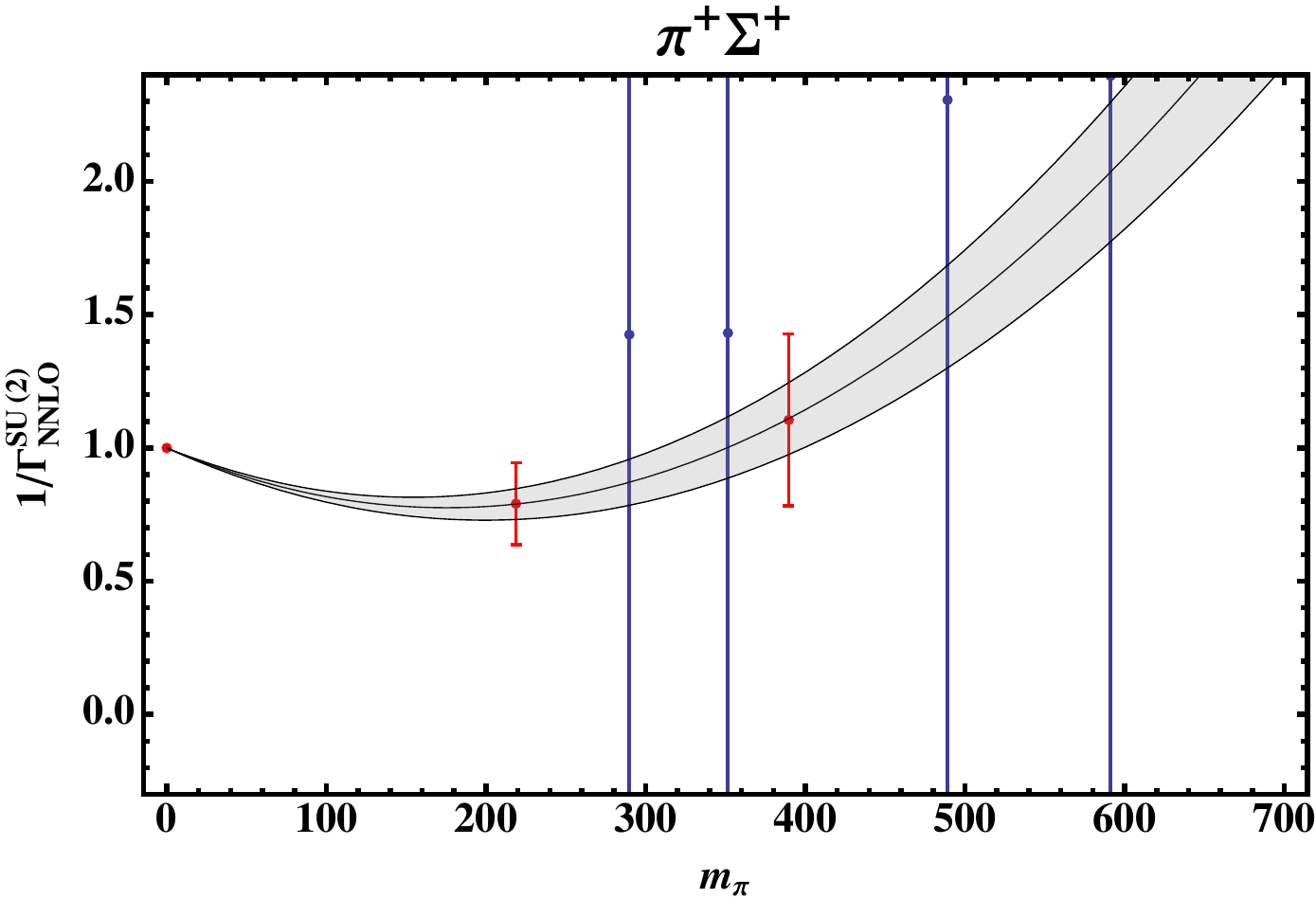} 
\caption{\label{fig:SigPi}$1/\Gamma_{\mathrm{NLO}}, 1/\Gamma_{\mathrm{NNLO}},$ and $1/\Gamma_{\mathrm{NNLO}}^{\mathrm{SU(2)}}$ (\Eq{Gamma}) for the $\pi^{+}\Sigma^{+}$ system as a function of the pion mass, for a representative value of the NNLO scale, $\mu$. Red points were calculated in this work, while blue are the data from \cite{Torok:2009dg}, shifted to take into account the effective range correction. The gray band represents the fit to the data using the rhs of \Eq{Gamma}, with all statistical and systematic errors included.}
\end{figure}

The results, including the shifted data, are shown in \Figs{SigPi}{pKnK}. We choose to investigate $1/\Gamma$ rather than $\Gamma$ because the effective range shift is large and of opposite sign to the inverse scattering length, thus, $\Gamma$ becomes an inverse of the difference between two noisy quantities which nearly cancel, causing the uncertainty on $\Gamma$ to become arbitrarily large. We expand the rhs of $1/\Gamma$ from \Eq{Gamma} to a given order in $m_{\pi}$ for the NLO and NNLO fits, also shown in these Figures. 

\begin{figure}
\includegraphics[width=0.48\linewidth]{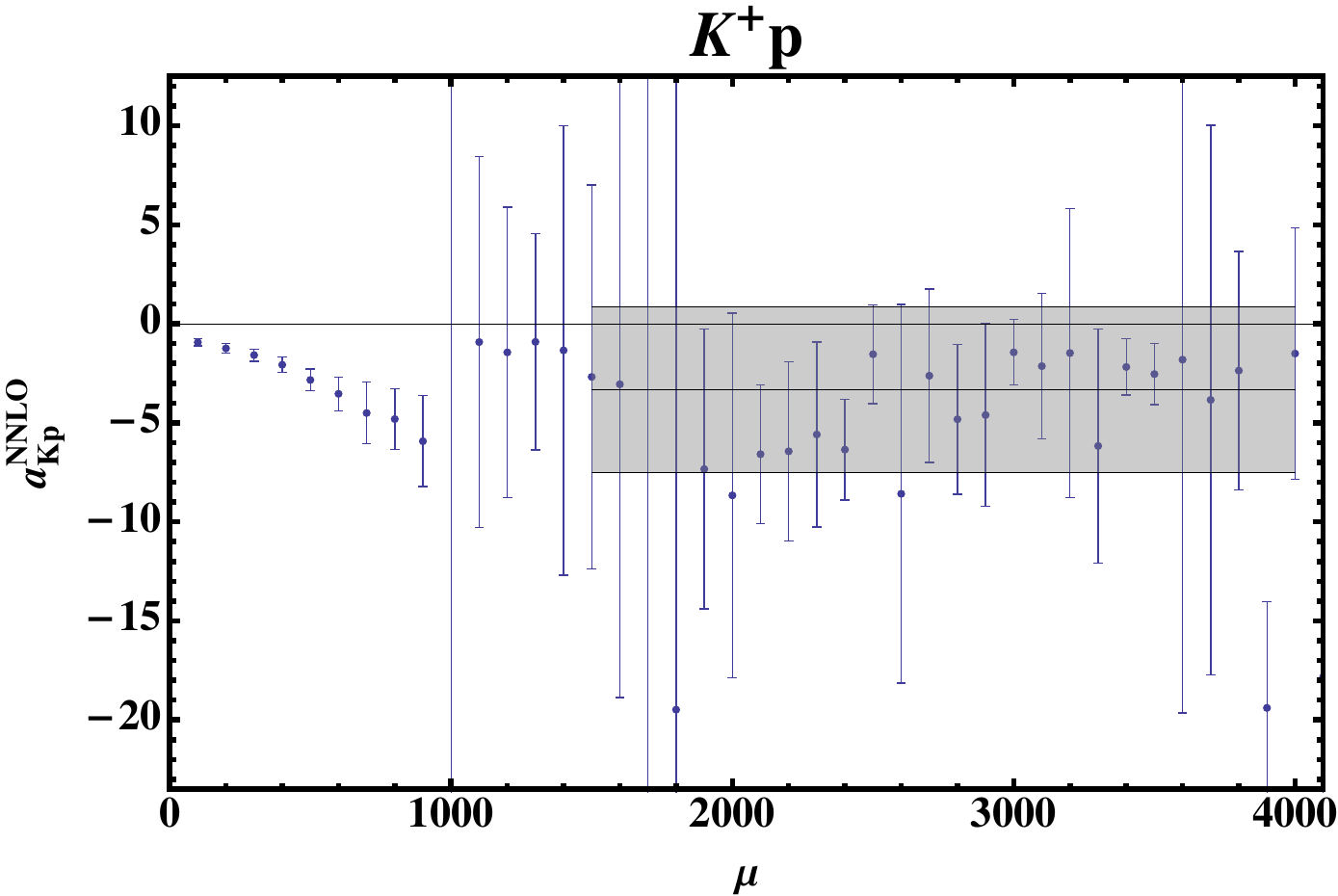}\hspace{1mm}
\includegraphics[width=0.48\linewidth]{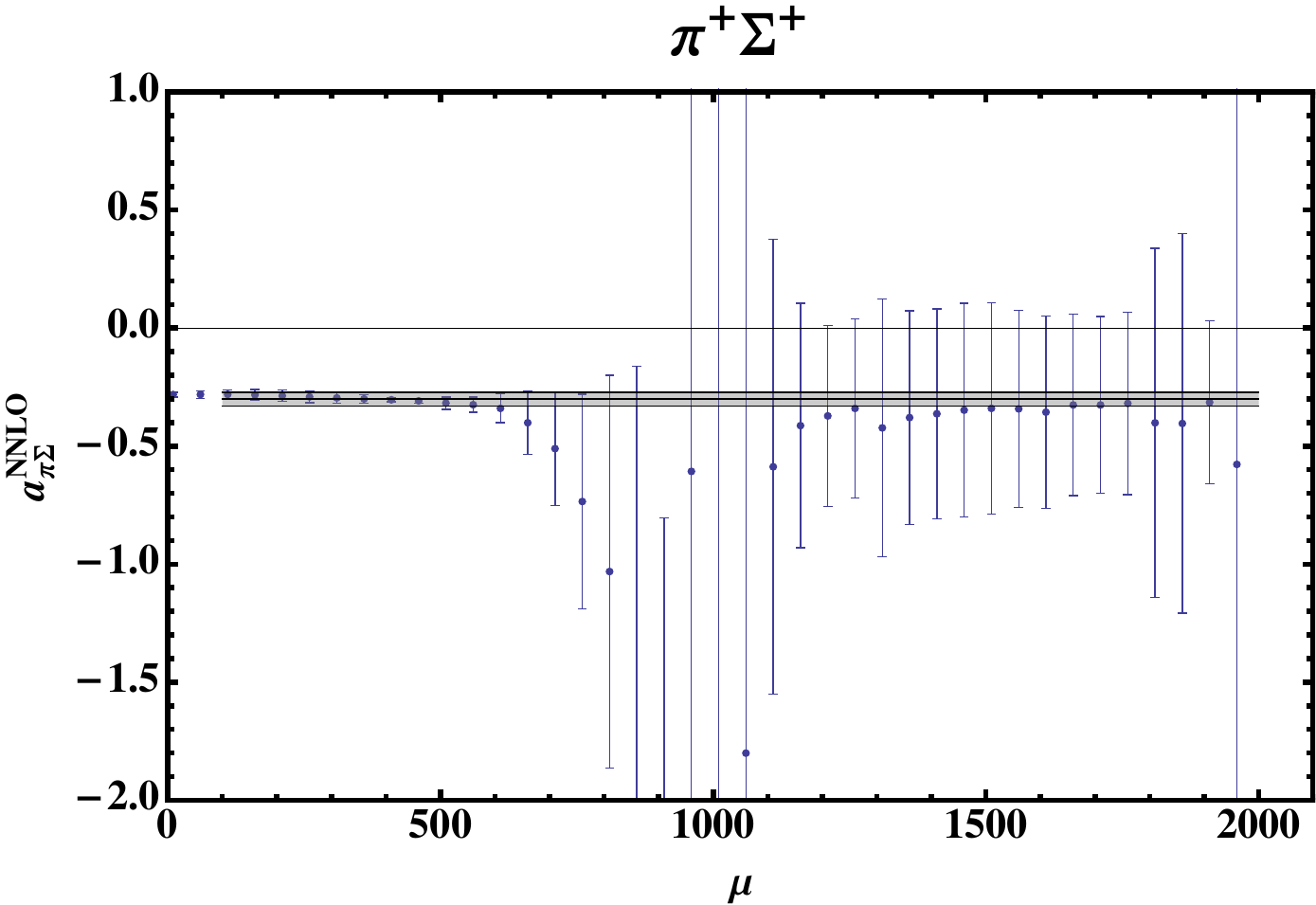} \\

\vspace{2mm}

\includegraphics[width=0.48\linewidth]{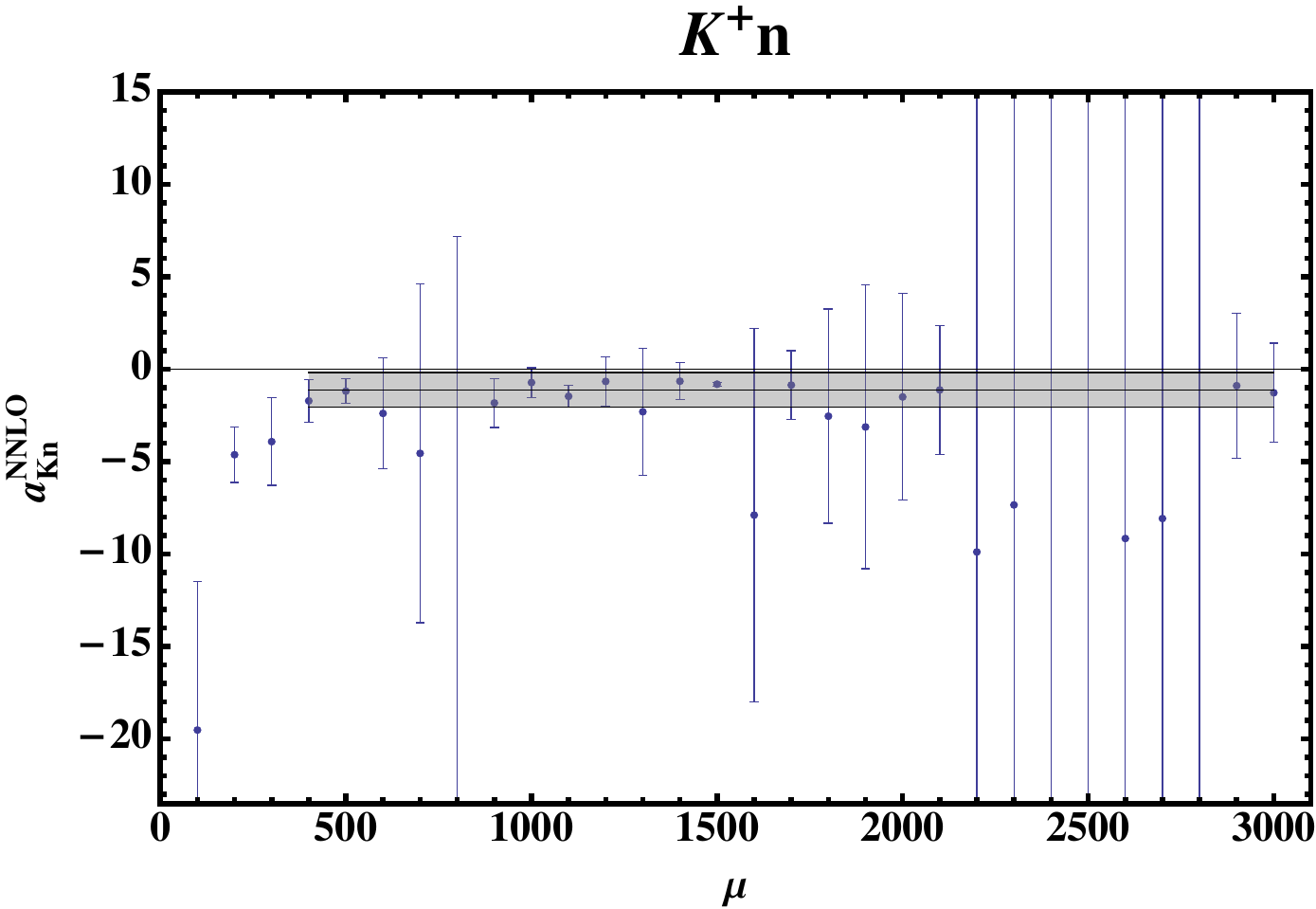} \hspace{1mm}
\includegraphics[width=0.48\linewidth]{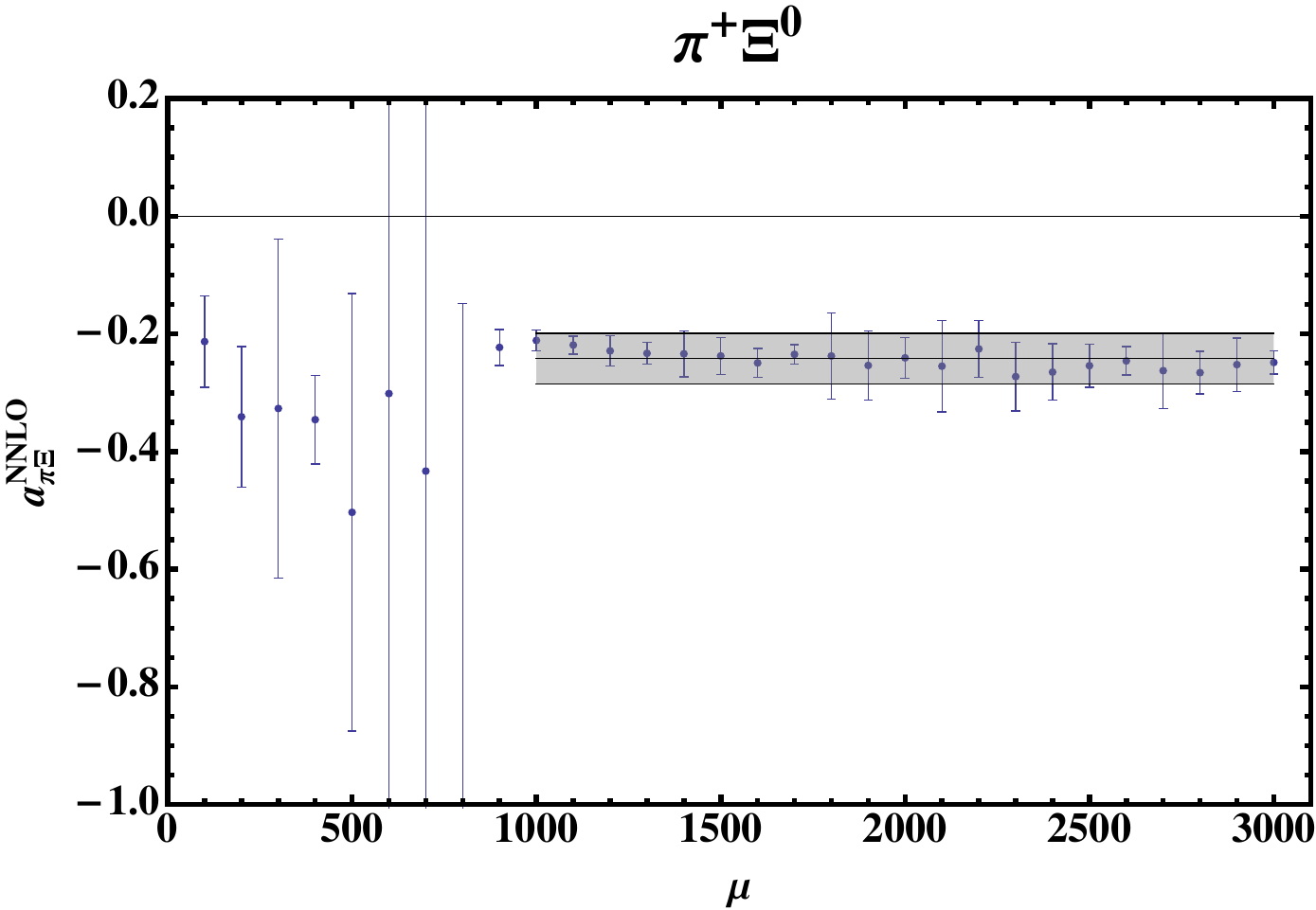}
\caption{\label{fig:mudependence}$\mu$-dependence of the scattering lengths at physical pion mass resulting from fits to the NNLO expressions (\Eq{Gamma}). The gray bands represent fits to the constant regions.}
\end{figure}

As noted in Ref.~\cite{Torok:2009dg}, at $\mu \sim \Lambda_{\chi}$ the NNLO shift to the data can be large (and is of opposite sign to NLO). This creates another noisy cancelation in $1/\Gamma$. We choose to scan the NNLO results over a wide range of $\mu$ and determine $a^{NNLO}$ from a fit to this set of data. This is demonstrated in \Fig{mudependence}. Note that the physical results become quickly $\mu$-independent as we raise $\mu$, and remains independent of the renormalization scale over a large range. This may be some indication that the next order in chiral perturbation theory is small.

\begin{figure}
\includegraphics[width=0.48\linewidth]{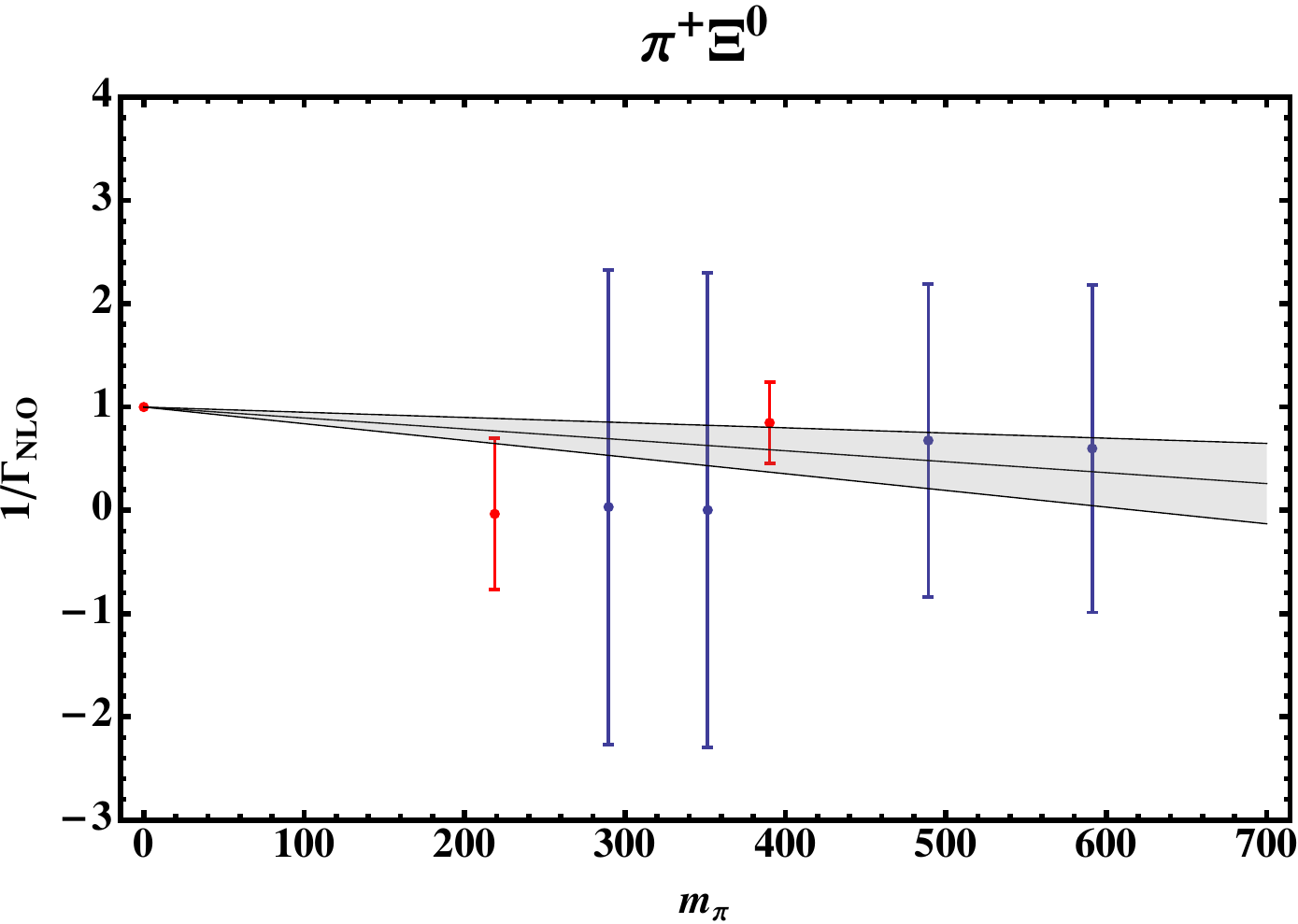}\hspace{1mm}
\includegraphics[width=0.48\linewidth]{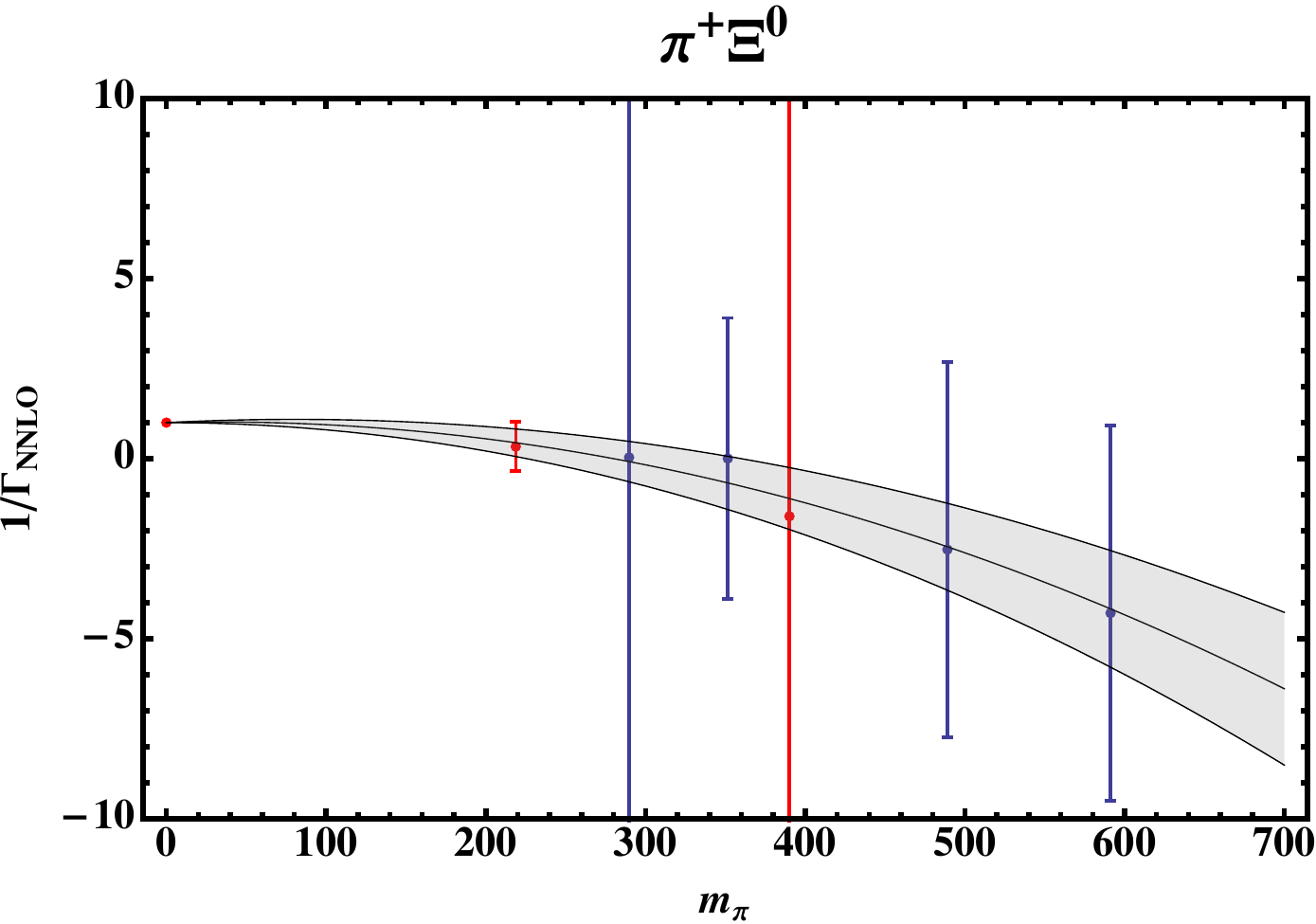} \\

\vspace{2mm}

\includegraphics[width=0.48\linewidth]{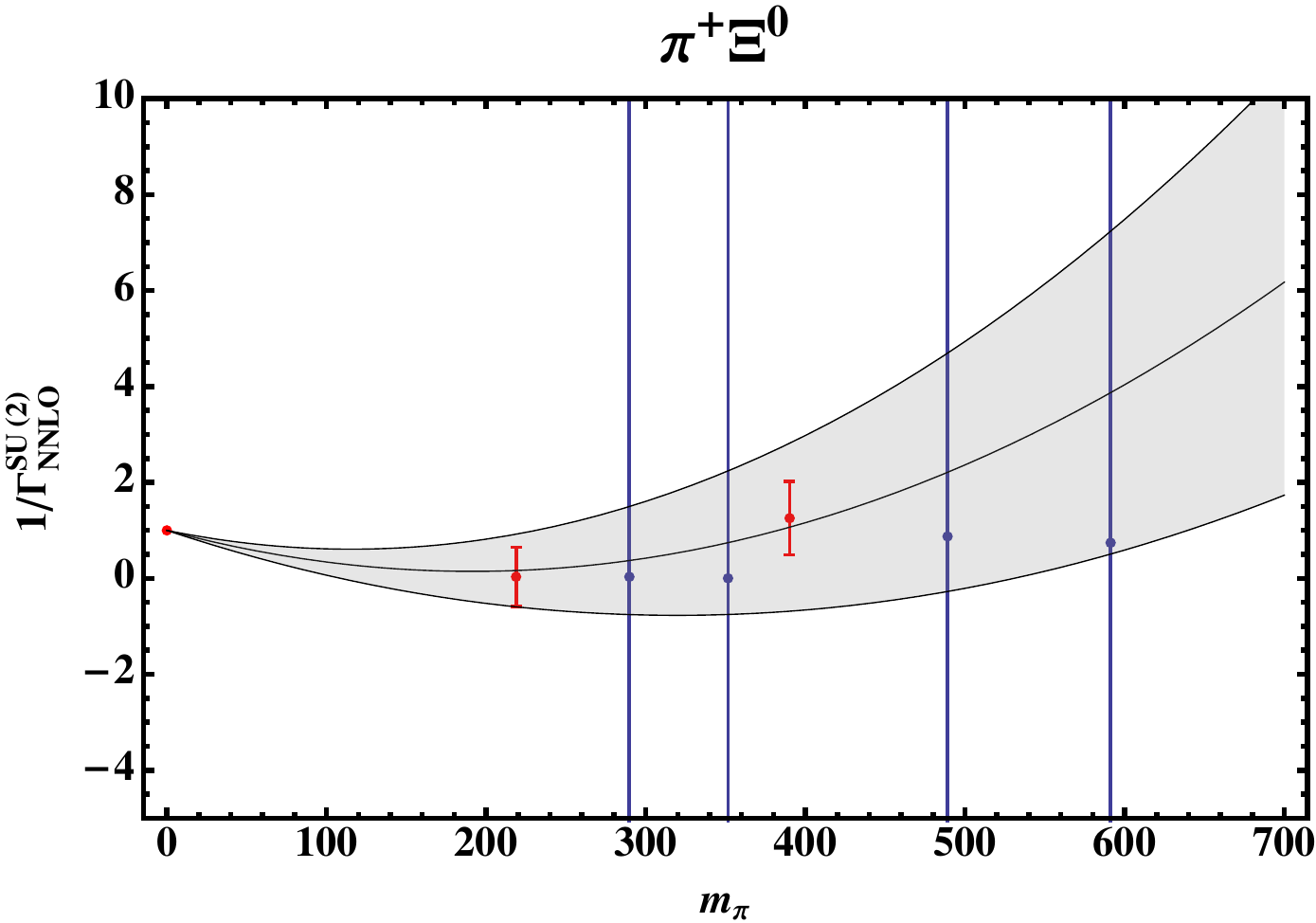} 
\caption{\label{fig:XiPi}$1/\Gamma_{\mathrm{NLO}}, 1/\Gamma_{\mathrm{NNLO}},$ and $1/\Gamma_{\mathrm{NNLO}}^{\mathrm{SU(2)}}$ (\Eq{Gamma}) for the $\pi^{+}\Xi^0$ system as a function of the pion mass, for a representative value of the NNLO scale, $\mu$. Red points were calculated in this work, while blue are the data from \cite{Torok:2009dg}, shifted to take into account the effective range correction. The gray band represents the fit to the data using the rhs of \Eq{Gamma}, with all statistical and systematic errors included.}
\end{figure}

\begin{figure}
\includegraphics[width=0.48\linewidth]{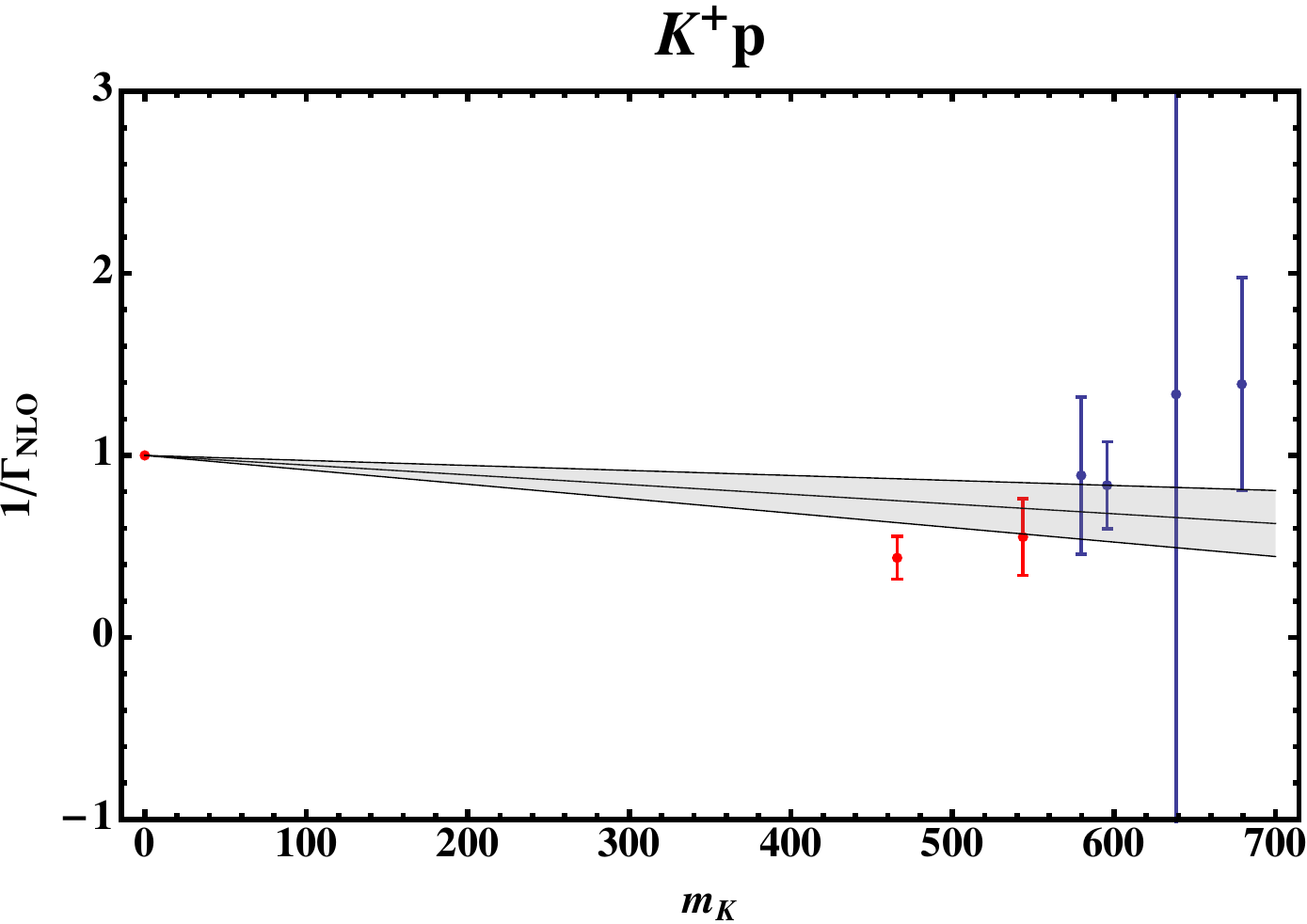}\hspace{1mm}
\includegraphics[width=0.48\linewidth]{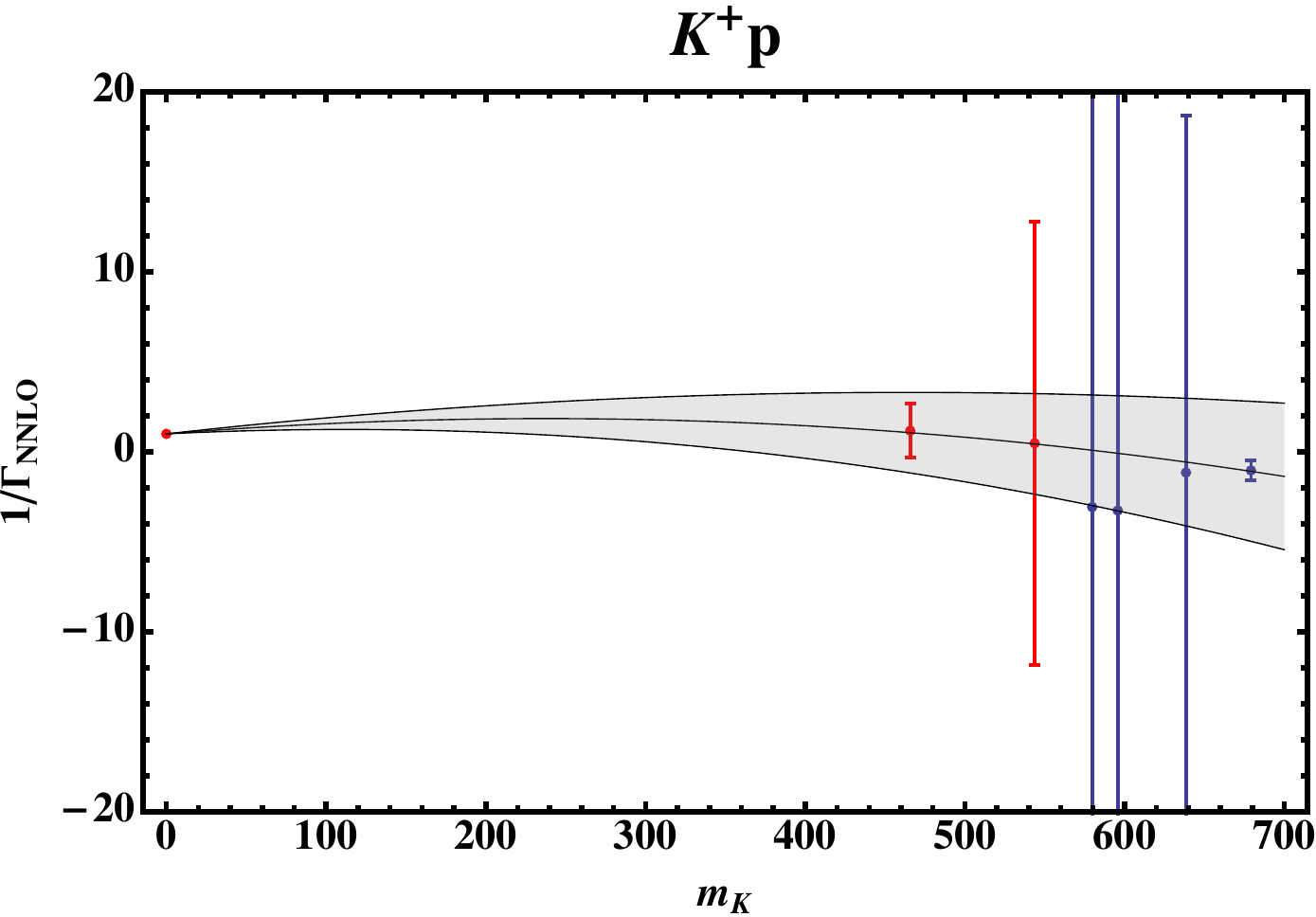} \\

\vspace{2mm}

\includegraphics[width=0.48\linewidth]{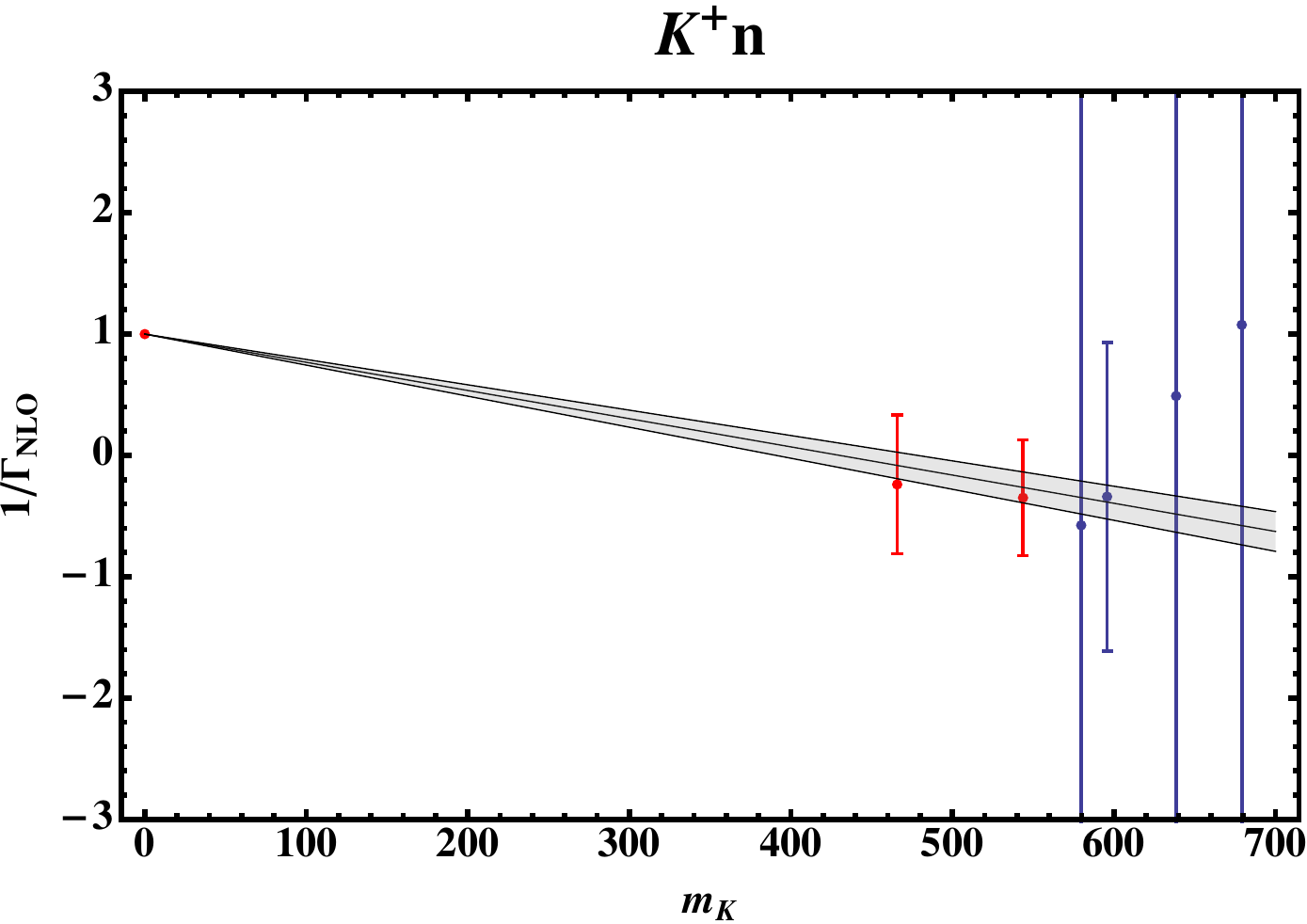} \hspace{1mm}
\includegraphics[width=0.48\linewidth]{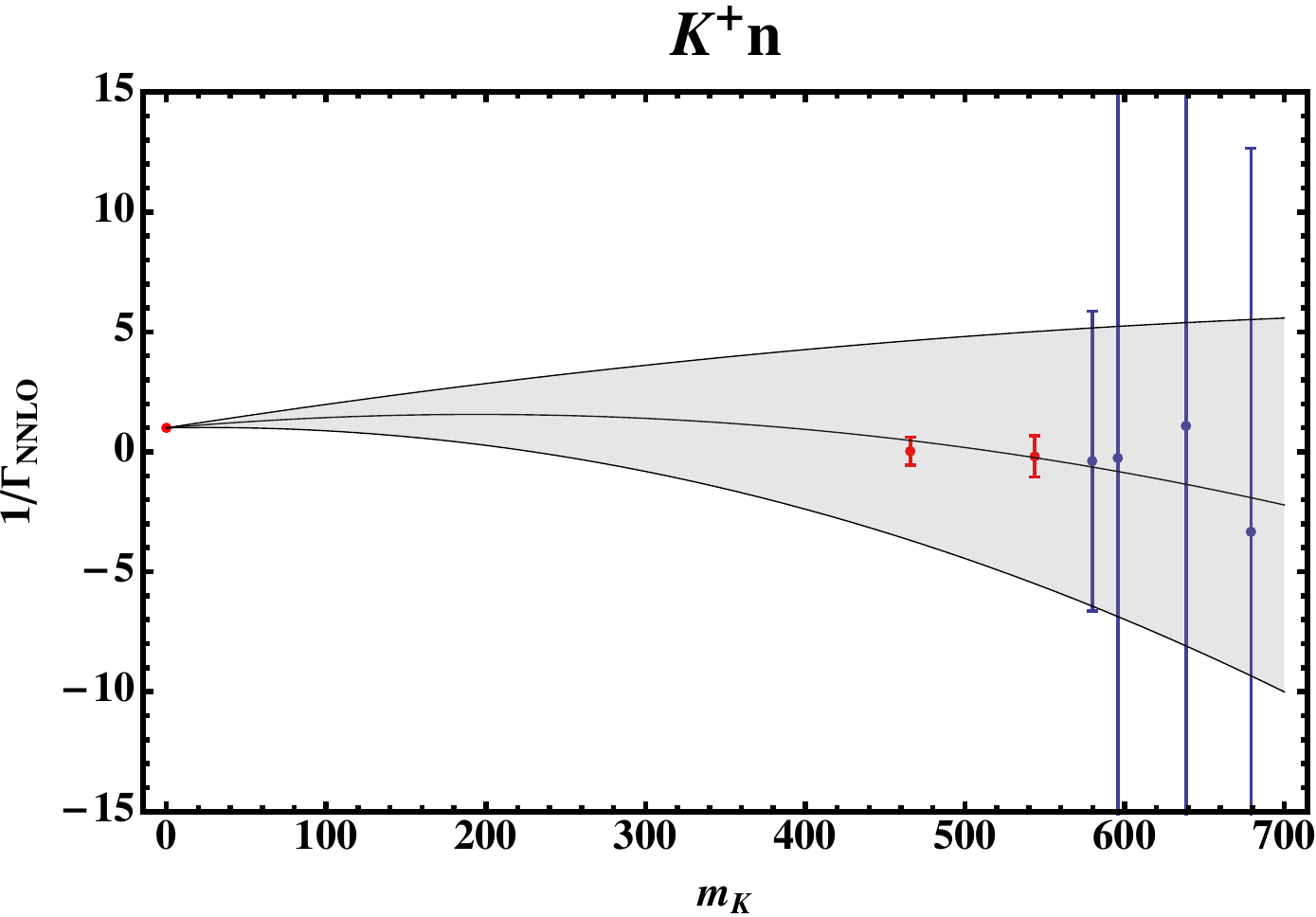}
\caption{\label{fig:pKnK}$1/\Gamma_{\mathrm{NLO}}$ and $1/\Gamma_{\mathrm{NNLO}}$ (\Eq{Gamma}) for the $K^{+}p$ (upper plots) and $K^{+}n$ (lower plots) systems as a function of the kaon mass, for a representative value of the NNLO scale, $\mu$. Red points were calculated in this work, while blue are the data from \cite{Torok:2009dg}, shifted to take into account the effective range correction. The gray band represents the fit to the data using the rhs of \Eq{Gamma}, with all statistical and systematic errors included.}
\end{figure}

\begin{figure}
\includegraphics[width=0.48\linewidth]{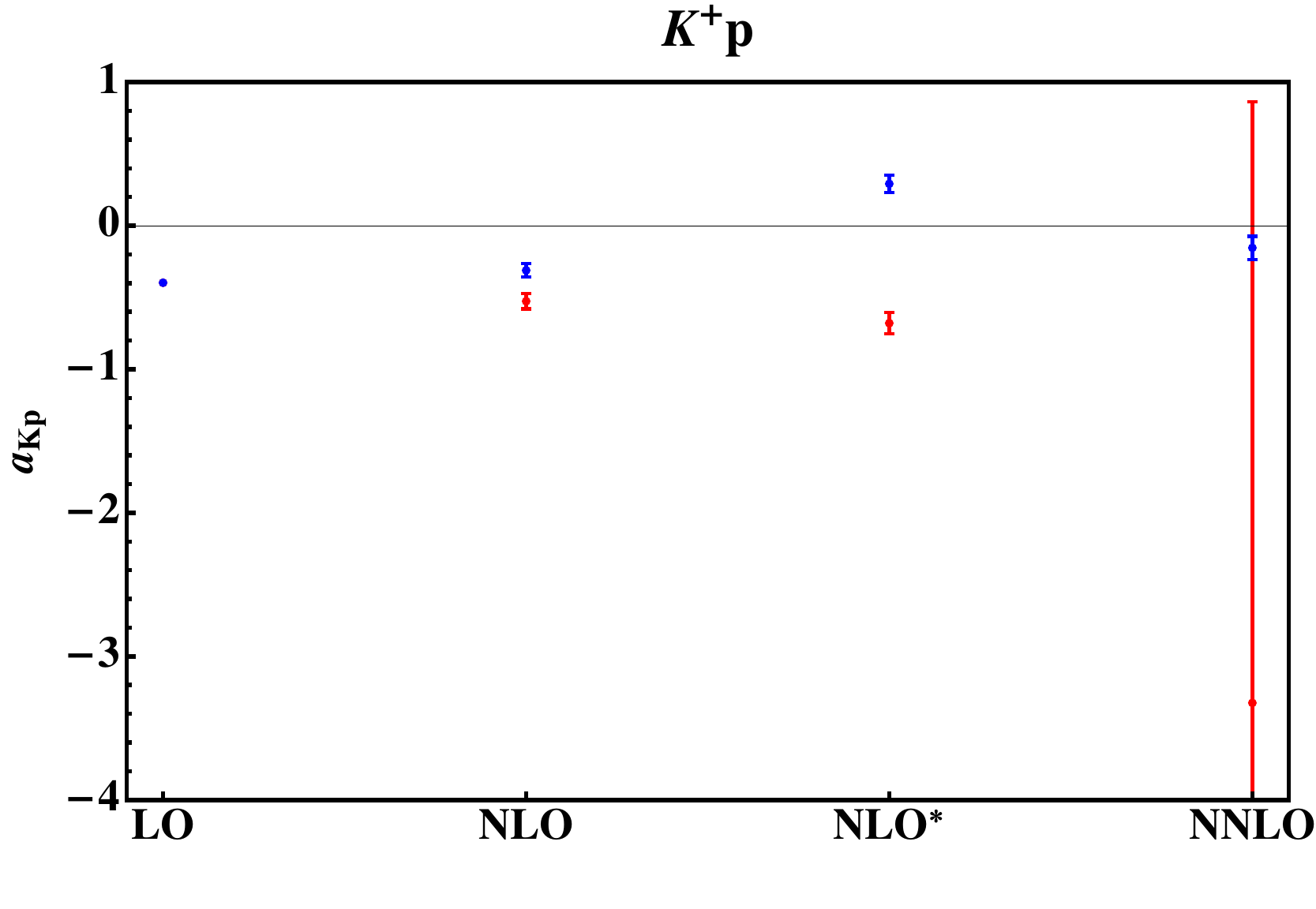}\hspace{1mm}
\includegraphics[width=0.48\linewidth]{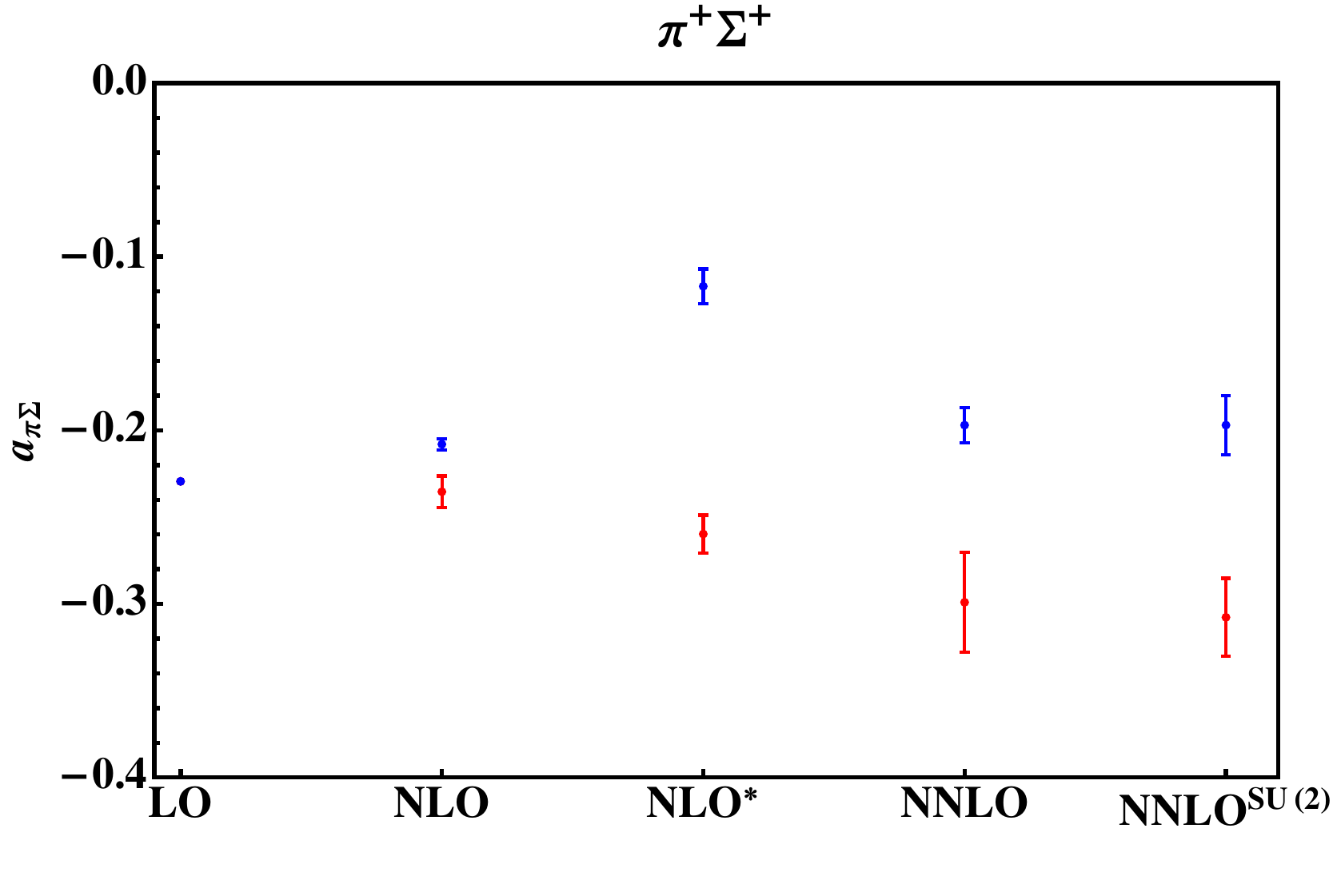} \\

\vspace{2mm}

\includegraphics[width=0.48\linewidth]{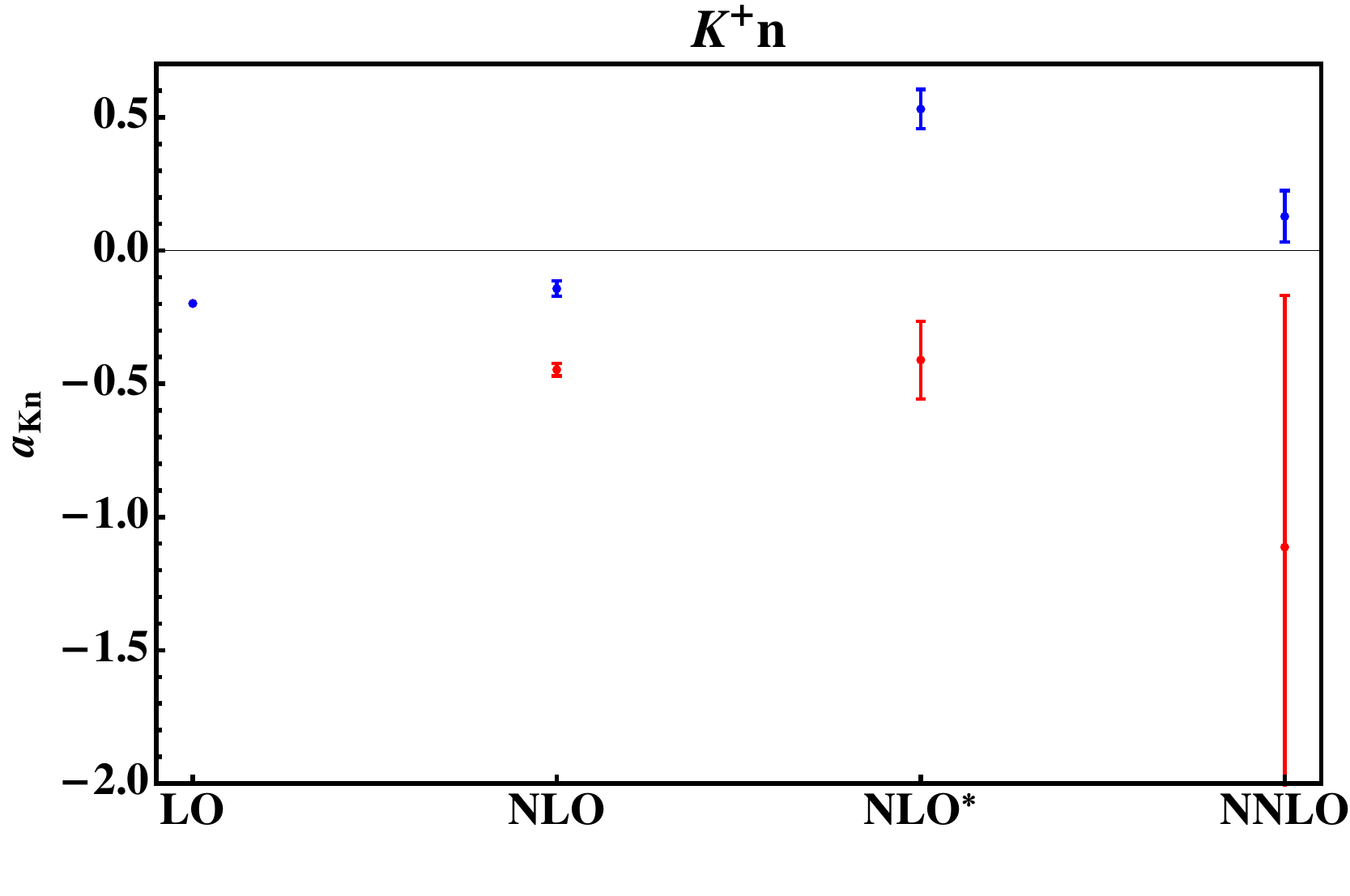} \hspace{1mm}
\includegraphics[width=0.48\linewidth]{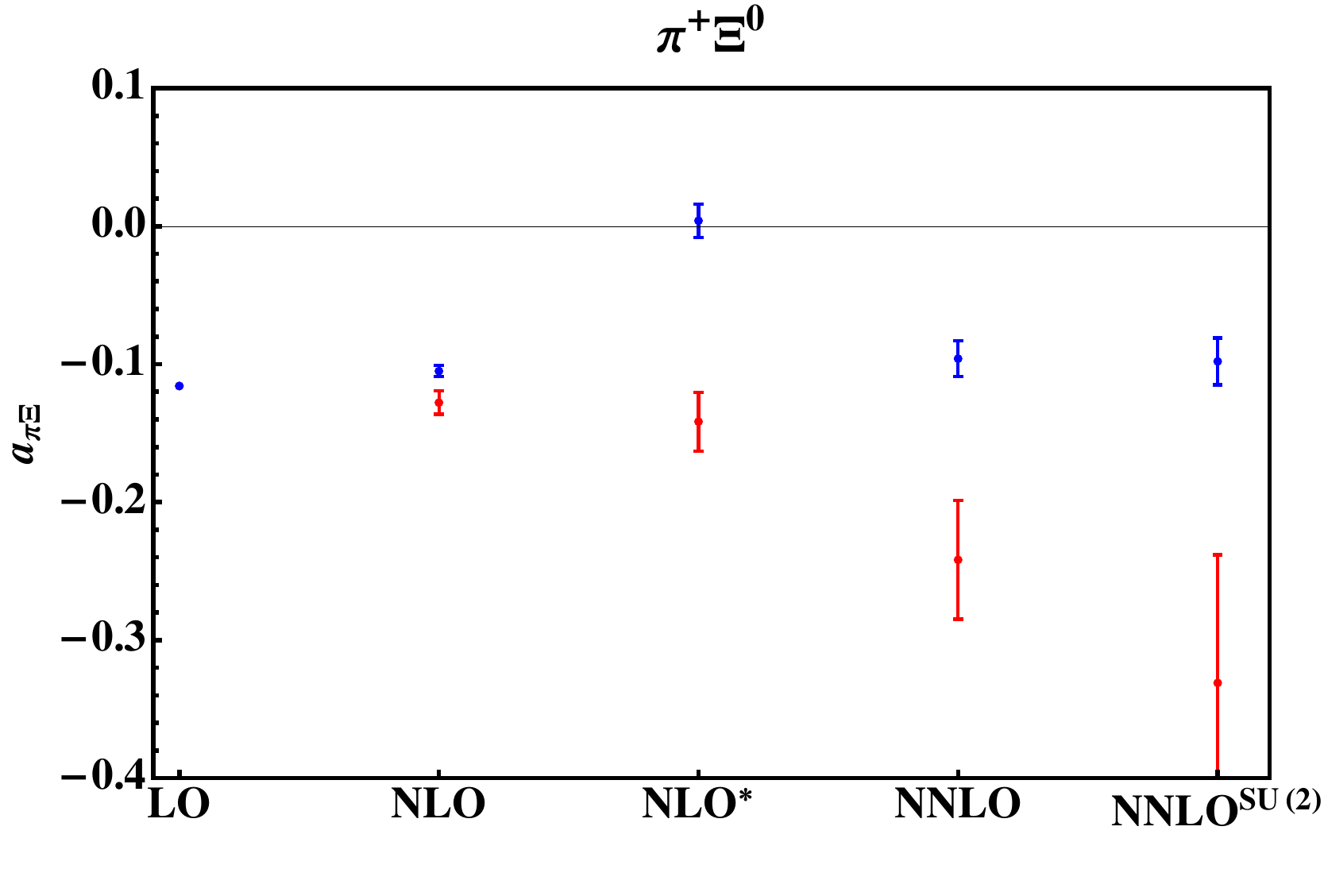}
\caption{\label{fig:scattlengths}Scattering lengths at the physical point resulting from Heavy Baryon $\chi$PT at LO, NLO, and NNLO. NLO$^{*}$ is the NLO result using the NNLO coefficients $C_1$ and $C_{01}$ from the pion-sigma and pion-xi systems. All results correspond to SU(3) $\chi$PT except for that labeled NNLO$^{SU(2)}$. Red points are those calculated in this work, while blue represent those stated in \cite{Torok:2009dg}. All error bars include statistical and systematic errors from the fit.}
\end{figure}

\begin{figure}
\includegraphics[width=0.48\linewidth]{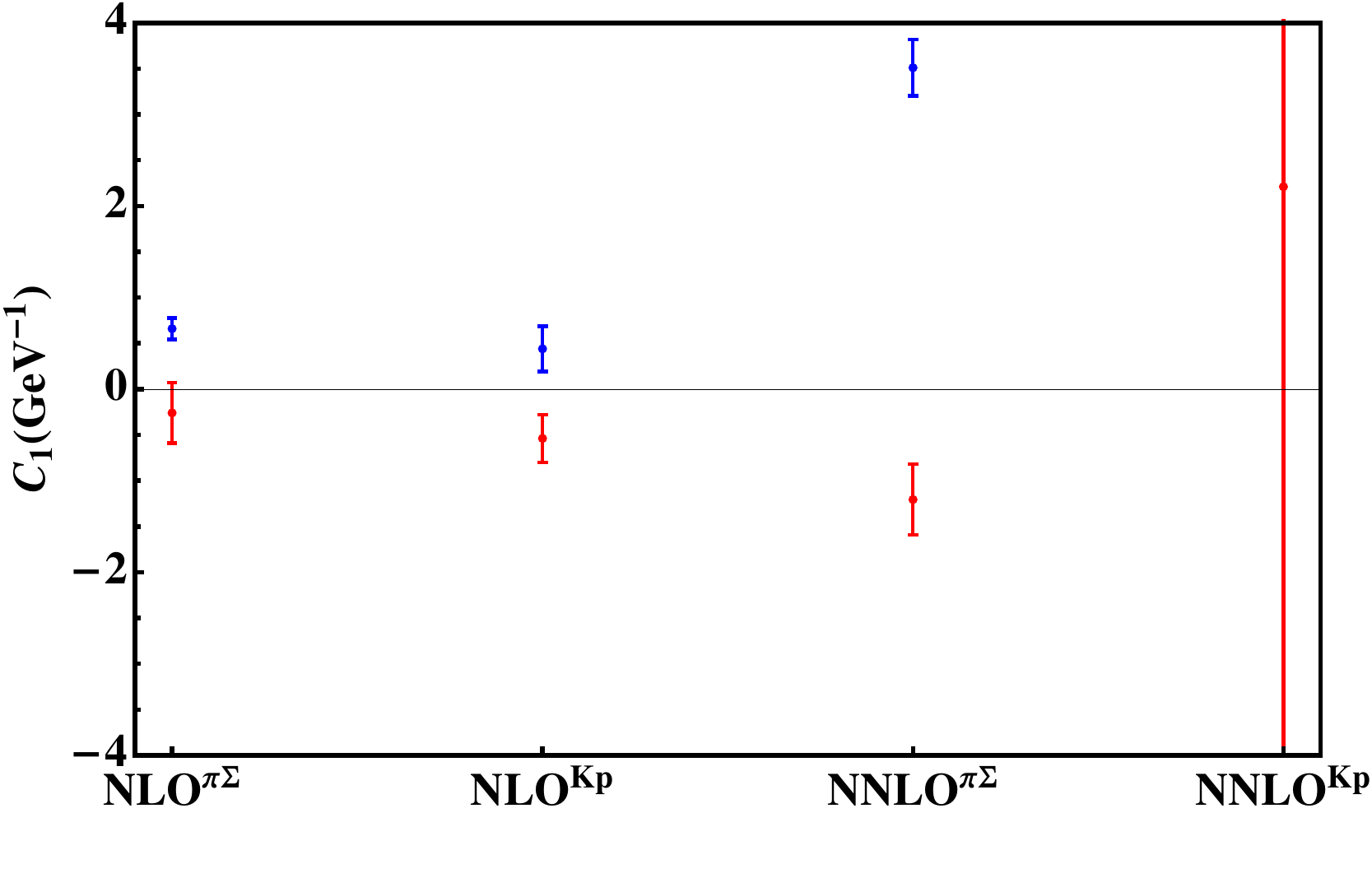}\hspace{1mm}
\includegraphics[width=0.48\linewidth]{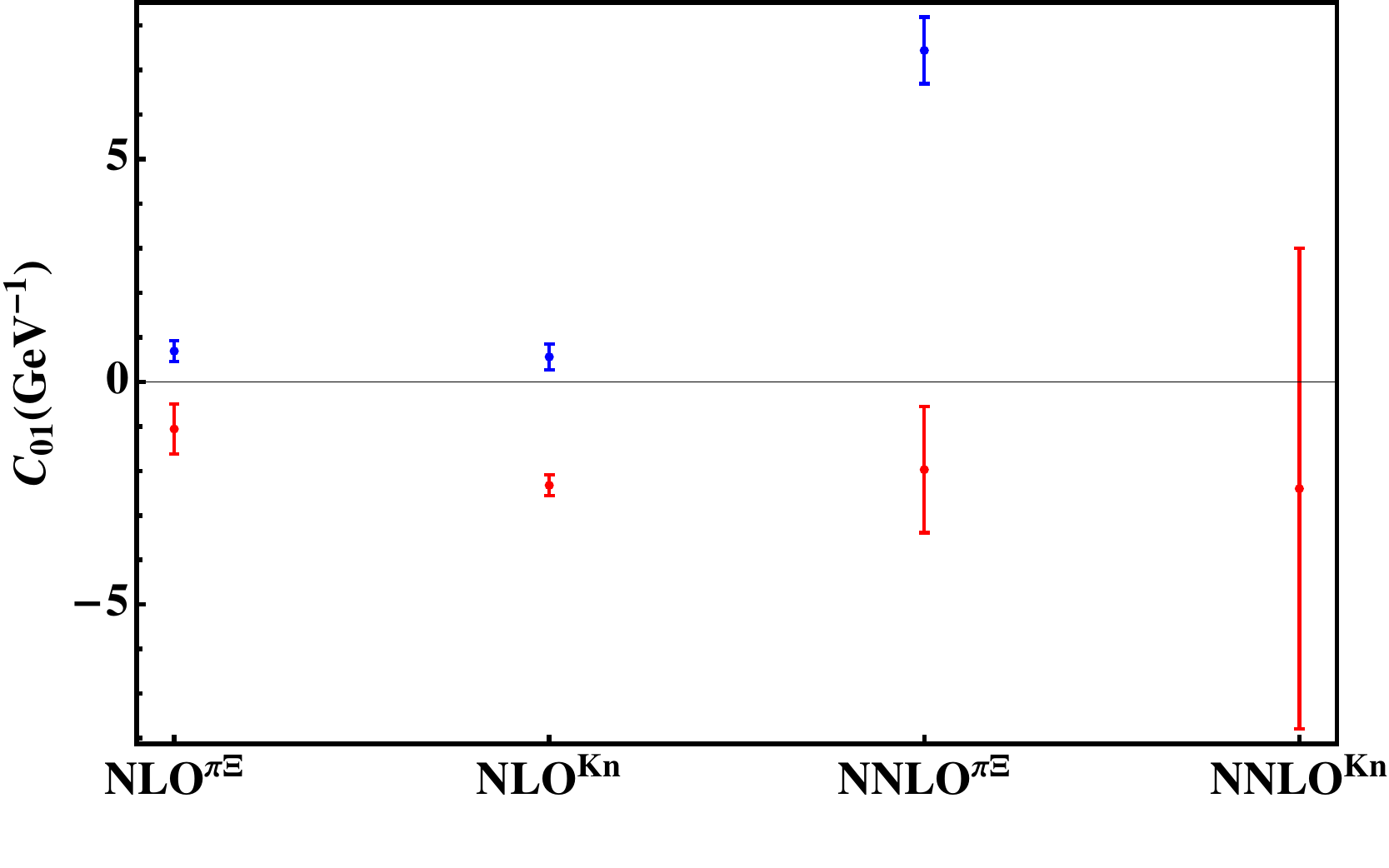} \\
\caption{\label{fig:LECs}Low energy constants $C_1$ and $C_{01}$ resulting from the fits to all four systems for NLO and NNLO SU(3) Heavy Baryon $\chi$PT. Red points are those calculated in this work, while blue represent those stated in \cite{Torok:2009dg}. All errors bars include statistical and systematic uncertainties.}
\end{figure}

In \Fig{scattlengths}, we present our results for the extrapolated scattering lengths to a given order in $\chi$PT, along with the unshifted results from Ref.~\cite{Torok:2009dg}. As in Ref.~\cite{Torok:2009dg} we define NLO$^{*}$ to be the resulting NLO value using the LECs from the NNLO fit to $\pi^{+}\Sigma^{+}, \pi^{+}\Xi^0$. Contrary to what was found in Ref.~\cite{Torok:2009dg} for the unshifted data, the SU(3) fits for the LECs appear to be fairly stable against the chiral corrections at the next order. This may be seen in \Fig{LECs}. We are unable to determine the NNLO result for the kaon systems due to noise. For the pion systems, the difference between the NNLO and NLO fits is somewhat large, particularly for $\pi^{+}\Xi^0$, although the correction is within the uncertainty of the results and is not significant. The NNLO results agree for the SU(2) and SU(3) extrapolations. Numerical values for the extrapolated scattering lengths and LECs are given in \Tab{extrap}. From the SU(2) extrapolation we find,
\beq
a^{(\pi\Sigma)}_{\mathrm{SU(2)}}(\mathrm{fm}) = -0.299(29) \ , \hspace{1mm} a^{(\pi\Xi)}_{\mathrm{SU(2)}}(\mathrm{fm}) = -0.242(43) \ , 
\eeq

\begin{table}
\centering
\caption{\label{tab:extrap}Scattering lengths extrapolated to the physical pion mass using SU(3) $\chi$PT at NLO and NNLO, as well as the NLO result using the NNLO coefficients $C_1$ and $C_{01}$ from the $\pi^{+}\Sigma^{+}$ and $\pi^{+}\Xi^0$ systems (NLO$^{*}$). Also included are the LECs, $C_1$ and $C_{01}$, that we have determined at NLO and NNLO.}
\begin{tabular}{|c|c|c|c|c|c|c|}
\hline
& \multicolumn{3}{c|}{$\pi^{+}\Sigma^{+}$} &\multicolumn{3}{c|}{$K^{+}p$}  \\
\hline
& NLO & NLO$^{*}$ & NNLO & NLO & NLO$^{*}$ & NNLO\\
\hline
$a(\mathrm{fm})$ &-0.2354(91) &-0.260(11) &-0.299(29) & -0.526(53)& -0.679(75)&-3.3(4.2)\\
$C_1(\mathrm{GeV}^{-1})$  & -0.26(33)& --- &-1.21(39)&-0.54(26)&---&2(11)\\
\hline
\end{tabular}
\begin{tabular}{|c|c|c|c|c|c|c|c|}
\hline
& \multicolumn{3}{c|}{$\pi^{+}\Xi^0$} &\multicolumn{3}{c|}{$K^{+}n $}  \\
\hline
& NLO &  NLO$^{*}$ & NNLO&NLO &  NLO$^{*}$& NNLO \\
\hline
$a (\mathrm{fm})$ & -0.1278(84)&-0.142(21) & -0.242(43)&-0.447(24) &-0.41(15)&-1.11(94)\\
$C_{01}(\mathrm{GeV}^{-1})$  &-1.06(56) & --- &-2.0(1.4)& -2.32(23)&--- &-2.4(54) \\
\hline
\end{tabular}
\end{table}

Given the stability of the LECs, the limited $\mu$-dependence (within errors), and the agreement between the SU(2) and SU(3) extrapolations, our results seem to indicate that $\chi$PT may be reliable for these quantities at the pion masses studied. However, because the uncertainties for the shifted data are much larger than the uncertainties from this work, all fits are generally dominated by these two points. Therefore, our statement about the stability of $\chi$PT is most likely limited to $m_{\pi}\lesssim 400$MeV.

\section{\label{sec:conclusions}Conclusions}

We have calculated the low-energy scattering phase shifts for four meson-baryon systems at a pion mass of $m_{\pi} \sim 390$ MeV, and determined the scattering lengths and effective ranges from this data. We find significant effective range contributions for scattering momenta corresponding to the three smallest volumes. These large effective ranges do not indicate a breakdown of the L\"uscher finite volume method, for which the relevant scale is the range of the interaction, $r_\mathrm{int}$. The effective ranges are, however, much larger than expected for a naturally tuned potential, where the range of the interaction and the effective range correspond to approximately the same scale. The leading long-range contribution to meson-baryon scattering comes from the exchange of two pions, so that $r_\mathrm{int} \sim 1/(2m_{\pi})$. While large effective ranges are indicative of fine-tuning of the potential, it is, however, fairly simple to tune a Yukawa potential, whose interaction range is set by the pion mass, to have arbitrarily large $r_0/a$ by varying the strength of the interaction within an order of magnitude of its ``natural" scale. 

Given the large effective ranges, we use these results to correct for range contributions in previous results, and combine these with our calculations at $m_{\pi}\sim 390$ and $m_{\pi} \sim 230$ MeV to perform a chiral extrapolation of the scattering lengths to physical pion mass. These differ significantly from the previous results \cite{Torok:2009dg}, which may be due to the effective range contributions and/or the poor convergence of HB$\chi$PT at heavier pion masses as noted in that work. For the lower pion masses used in this work, we find relative stability of the SU(3) HB$\chi$PT expansion, and agreement at NNLO with SU(2) HB$\chi$PT.

\begin{acknowledgments}
The authors would like to thank R. Brice\~{n}o, A. Walker-Loud, T. Kurth, and E. Berkowitz for helpful discussions. We thank the NPLQCD collaboration for the use of previously generated meson and baryon blocks. Portions of our calculations were performed at NERSC (a DOE Office of Science User Facility supported by the U.S. Department of Energy under Contract No. DE-AC02-05CH11231), and on USQCD supported facilities. WD was supported in part by the U.S. Department of Energy through Early Career Award DE-SC0010495. AN was supported in part by U.S. DOE grant No. DE-SC00046548. 

\end{acknowledgments}
\bibliography{Mesonsref}

\end{document}